\documentclass[10pt,twocolumn,letterpaper]{article}

\usepackage[pagenumbers]{cvpr}

\input{preamble}

\definecolor{cvprblue}{rgb}{0.21,0.49,0.74}
\usepackage[pagebackref,breaklinks,colorlinks,allcolors=cvprblue]{hyperref}
\hypersetup{
  pdftitle={3DSS: 3D Surface Splatting for Inverse Rendering},
  pdfauthor={Mae Younes, Adnane Boukhayma},
  pdfkeywords={differentiable rendering, surface splatting, inverse rendering, surfels, EWA splatting}
}

\def\confName{CVPR}
\def\confYear{2027}

\title{3DSS: 3D Surface Splatting for Inverse Rendering}

\author{Mae Younes \qquad Adnane Boukhayma\\
INRIA, University of Rennes, France
}

\begin{document}
\twocolumn[{%
\renewcommand\twocolumn[1][]{#1}%
\maketitle
\vspace{-22pt}
\begin{center}
    \centering
    \captionsetup{type=figure}
    \includegraphics[width=\textwidth]
    {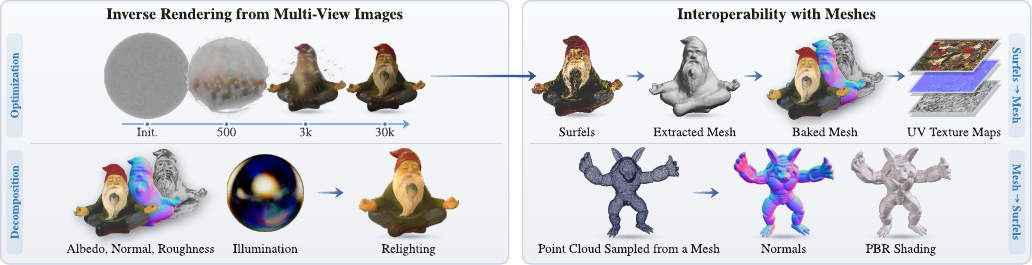}
    \captionof{figure}{\emph{3D Surface Splatting} (3DSS) is a differentiable surface splatting renderer for physically-based inverse rendering. Left: optimizing unstructured surfels jointly recovers geometry, materials, and illumination from multi-view images. Right: the optimized surfels convert into a mesh, into which their materials bake as vertex attributes and UV texture maps (top); conversely, 3DSS renders oriented point clouds sampled from triangle meshes with anti-aliased, physically-based shading (bottom).}
    \label{fig:teaser}
    \vspace{-4pt}
\end{center}%
}]
\begin{abstract}
We present \emph{3D Surface Splatting} (3DSS), the first differentiable surface splatting renderer for physically-based inverse rendering from multi-view images. Our central insight is that surface separation, which groups the surfels covering a pixel by the surface they sample, can be formulated with the reconstruction kernels themselves: two surfels sample the same surface when their kernels overlap in depth. Compositing each resulting surface layer with a coverage opacity derived from its accumulated Elliptical Weighted Average (EWA) weight yields anti-aliased silhouettes and informative visibility gradients at sparsely covered edges. With forward microfacet shading under co-optimized HDR environment lighting, 3DSS jointly recovers shape, spatially-varying BRDF, and illumination, and its oriented surfels convert directly into meshes by oriented-point surface reconstruction. On the real-world Stanford-ORB and the TensoIR Synthetic benchmarks, 3DSS outperforms state-of-the-art mesh-based and major comparable implicit and Gaussian-based methods in novel view synthesis and relighting, with geometry on par with the strongest baselines.
\end{abstract}

\section{Introduction}
\label{sec:intro}

Inverse rendering recovers geometry, materials, and illumination from photographs, and requires a renderer that is differentiable, including through visibility, which indicates whether a surface point is the first intersection along a camera ray or is occluded by closer geometry.
Visibility is binary: its derivative vanishes almost everywhere and is undefined exactly at silhouettes and occlusion boundaries, where the geometry most needs guidance.
Physically-based differentiable path tracers handle this in a principled way, by edge sampling or reparameterization~\cite{li2018differentiable,loubet2019reparameterizing,zhang2020path}, at a cost that weighs on long optimizations.
Most multi-view methods therefore rasterize or splat, and how they handle visibility depends on the scene representation.

Mesh rasterizers such as \textsc{nvdiffrast}~\cite{laine2020modular}, combined with the tetrahedral grid of DMTet~\cite{shen2021deep}, recover explicit surfaces, materials, and lighting~\cite{Munkberg_2022_CVPR,hasselgren2022shape}.
A mesh whose topology evolves, however, needs careful regularization against degenerate elements, self-intersections, and broken manifolds, failures that worsen as the grid is refined.

Volume rendering of neural fields~\cite{mildenhall2021nerf,wang2021neus} or 3D Gaussians~\cite{kerbl2023gaussian} needs no connectivity and excels at novel view synthesis, but it composites \emph{along the ray} rather than \emph{at the surface}.
In inverse rendering~\cite{liang2024gs,sun2025svg,gu2024IRGS}, this blends material attributes and normals into values that belong to no single surface, undermining the decomposition itself.

An alternative has remained absent from differentiable inverse rendering: Elliptical Weighted Average (EWA) surface splatting~\cite{zwicker2001surface}, in which oriented surfels (surface elements) with Gaussian reconstruction kernels in their tangent planes reconstruct a continuous, band-limited surface signal.
It combines the connectivity-free flexibility of points with the surface-based image formation of meshes: each surfel samples the surface itself, so each pixel resolves to an opaque surface point.

The obstacle is that a surface \emph{emerges} from overlapping kernels, not from any single primitive, so visibility cannot be localized to one surfel as it can to a triangle edge.
The renderer must first solve \emph{surface separation}: decide, at each pixel, which surfels sample the same surface and must be averaged, and which lie on another surface in front or behind.
Classical splatters~\cite{zwicker2001surface,weyrich2007hardware} keep only the frontmost surface through a ternary depth test, a hard decision that is incompatible with gradient-based optimization and precludes correct anti-aliasing across overlapping surfaces.

We present \emph{3D Surface Splatting} (3DSS), the first differentiable surface splatting renderer for physically-based inverse rendering, which jointly recovers geometry, materials, and illumination from multi-view images (\cref{fig:teaser}).
Our central insight is that surface separation can be expressed with the reconstruction kernels themselves: two surfels sample the same surface at a pixel when their kernels overlap in depth.
Three contributions make 3DSS surface-based by construction, anti-aliased, and differentiable through visibility:
\begin{itemize}[nosep,leftmargin=*]
  \item \textbf{Depth-interval grouping}, which separates multiple surface layers per pixel in one rasterization pass (\cref{sec:depth_grouping}); the same rule resolves visibility along the secondary rays of our Monte Carlo shading (\cref{sec:mc_shading}).
  \item \textbf{Coverage-based compositing}, in which each layer's opacity follows from its accumulated EWA weight and layers are composited front to back with the over operator~\cite{porter1984compositing}, yielding anti-aliased silhouettes and informative visibility gradients at sparsely covered edges (\cref{sec:compositing}).
  \item \textbf{A center-precomputed MIP filter} that matches the anti-aliasing of the object-space MIP filter of AA-2DGS~\cite{younes2025anti} at a fraction of its overhead (\cref{sec:mip_filtering}).
\end{itemize}

As oriented surface samples, our surfels convert directly into meshes by oriented-point surface reconstruction~\cite{huang2023nksr}, onto which their materials bake as vertex attributes or UV textures (\cref{fig:teaser} and \cref{sec:supp_meshes}).
We evaluate 3DSS on the real-world Stanford-ORB~\cite{kuang2023stanfordorb}, with laser scans and measured lighting, and on TensoIR Synthetic~\cite{jin2023tensoir}, with ground-truth materials and thin structures.
Against mesh-based, implicit, and Gaussian-splatting baselines, notably IRGS~\cite{gu2024IRGS}, built on 2D Gaussian splatting~\cite{huang20242d}, and the mesh-based NVDiffRecMC~\cite{hasselgren2022shape}, it achieves the best novel view synthesis and relighting on both, with geometry on par with the strongest baselines.

\section{Related Work}
\label{sec:related}

\paragraph{Surface splatting.}
\label{sec:rw_ewa}
\label{sec:rw_dss}
Surface splatting~\cite{zwicker2001surface} adapted the EWA resampling filter~\cite{heckbert1989fundamentals,greene2007creating} to point-based rendering~\cite{levoy1985use}, resolving surface separation with a fixed-threshold ternary depth test (extended z-buffering~\cite{krivanek2003representing}); per-splat depth ranges~\cite{levoy1985use,rasanen2002surface} instead merge overlapping fragments.
Weyrich et al.~\cite{weyrich2007hardware} parameterize the ternary test by each surfel's depth extent and intersect rays with surfels in object space, avoiding the numerical instabilities of screen-space conics~\cite{zwicker2004perspective}.
Most of these methods keep only the frontmost surface per pixel, precluding edge anti-aliasing across overlapping surfaces, and their hard depth tests are not differentiable; our depth-interval grouping extends depth ranges to any number of surface layers in one rasterization pass (\cref{sec:depth_grouping}).
DSS~\cite{yifan2019differentiable} made EWA splatting differentiable for geometry processing, approximating visibility gradients by hypothetically toggling each surfel, which requires caching and per-pixel re-evaluation.
We instead handle visibility per surface layer, through its coverage (\cref{sec:coverage}), and target physically-based inverse rendering (\cref{sec:forward_shading}).

\paragraph{Differentiable mesh rendering.}
\label{sec:rw_rast}
The central challenge of differentiable rendering is to produce gradients through the visibility function, which switches discontinuously between visible and occluded at silhouettes and occlusion boundaries.
SoftRas~\cite{liu2019soft} blurs triangle coverage at the cost of image fidelity; DIB-R~\cite{chen2019learning} derives a soft alpha from edge distances but stays aliased.
\mbox{\textsc{nvdiffrast}}~\cite{laine2020modular} keeps the image crisp and obtains visibility gradients from an analytic anti-aliasing pass across silhouette edges, which requires triangle connectivity, acts only on axis-aligned pixel pairs, and can miss edges under fine tessellation.
Differentiable path tracers~\cite{li2018differentiable,loubet2019reparameterizing,bangaru2020unbiased,zhang2020path,nimier2019mitsuba} treat visibility rigorously but are orders of magnitude slower than rasterization.
\mbox{NVDiffRec}~\cite{Munkberg_2022_CVPR} couples \mbox{\textsc{nvdiffrast}} with DMTet~\cite{shen2021deep} and split-sum image-based lighting (IBL)~\cite{karis2013real}, and \mbox{NVDiffRecMC}~\cite{hasselgren2022shape}, our representative mesh baseline, replaces the split sum with ray-traced direct illumination.

\paragraph{Volumetric inverse rendering.}
\label{sec:rw_nerf}
\label{sec:rw_3dgs}
Neural Radiance Field (NeRF)~\cite{mildenhall2021nerf} represents scenes as continuous volumetric radiance fields parameterized by MLPs, queried along camera rays and composited via volume rendering.
Subsequent work has improved geometric fidelity via signed distance functions (SDF)~\cite{yariv2020multiview,wang2021neus,yariv2021volume}.
Physically-based decompositions~\cite{zhang2021nerfactor,zhang2021physg,boss2021nerd,zhang2022invrender,liu2023nero,jin2023tensoir,sun2023neural} of NeRF and SDF-based fields mostly inherit the cost of ray marching.
3D Gaussian Splatting (3DGS)~\cite{kerbl2023gaussian} instead alpha-composites anisotropic 3D Gaussians with learnable opacities, and 2D Gaussian Splatting (2DGS)~\cite{huang20242d} flattens them into oriented planar disks intersected with rays in object space~\cite{weyrich2007hardware}.
Gaussians with physically-based attributes model reflective appearance~\cite{jiang2024gaussianshader,ye20243d,yao2024reflective,zhang2025ref,younes2025texturesplat} or recover materials and illumination~\cite{gao2023relightable,shi2023gir,liang2024gs,zhou2025rtr,sun2025svg,gu2024IRGS,jiang2025radiositygs}.
GS-IR~\cite{liang2024gs} shades 3DGS with split-sum IBL and bakes occlusion into spherical-harmonic probes.
IRGS~\cite{gu2024IRGS}, our primary baseline, builds on 2DGS and evaluates the rendering equation by Monte Carlo sampling, alpha-blending 2D Gaussians along traced secondary rays.
We share the planar disks and object-space intersection of 2DGS and IRGS but not their volumetric image formation: our surfels have no learnable opacity and are averaged within a surface layer, not alpha-blended.
Mip-Splatting~\cite{yu2024mip} prefilters 3DGS in screen space, whereas AA-2DGS~\cite{younes2025anti} maps the prefilter into object space through the Jacobian of the ray--surfel intersection; we adopt this object-space MIP filter, with the Jacobian precomputed at surfel centers (\cref{sec:mip_filtering}).

\section{Method}
\label{sec:method}

\subsection{Surfels and Rasterization}
\label{sec:surfel_primitive}
3DSS renders an unstructured set of surfels by EWA surface splatting~\cite{zwicker2001surface}.
A surfel has a center $\mathbf{c}$, two tangent vectors $\mathbf{t}_u, \mathbf{t}_v$ that span its tangent plane and encode its orientation and anisotropic scale, an albedo $\boldsymbol{\rho}$, a roughness $r$, a metallic $m$, and a degree-3 spherical-harmonics (SH) radiance function (\cref{sec:forward_shading}).
Its normal is $\mathbf{n} = \mathbf{t}_u \times \mathbf{t}_v / \|\mathbf{t}_u \times \mathbf{t}_v\|$, and its \emph{object space} maps coordinates $(u,v)$ to $\mathbf{c} + u\,\mathbf{t}_u + v\,\mathbf{t}_v$; there, the reconstruction kernel is a unit-variance isotropic Gaussian~\cite{weyrich2007hardware}, so that all anisotropy is carried by the tangent vectors, which we parameterize by a unit quaternion and two scales of bounded ratio (details in \cref{sec:parameterization}).

The T-matrix of Weyrich et al.~\cite{weyrich2007hardware}, $\mathbf{T} = \mathbf{P}\mathbf{S}$, which composes the object-to-camera mapping $\mathbf{S} = [\,\mathbf{t}'_u\;\mathbf{t}'_v\;\mathbf{0}\;\mathbf{c}'\,]$ (primes mark camera space) with the projection $\mathbf{P}$, gives both the screen-space bounding box of a surfel and, at each pixel, the object-space coordinates $(u,v)$ of the ray--surfel intersection (details in \cref{sec:supp_tmatrix}).
Back-facing surfels are culled.
A front-facing surfel contributes the weight $w$ of its unnormalized kernel, with cutoff $r_{\text{cut}} = 3$, and an attribute $\mathbf{a}$ is reconstructed by Shepard normalization over the surfels of one \emph{surface layer}, \ie, those that sample the same surface:
\begin{gather}
  w = \exp\!\bigl(-\tfrac{1}{2}\,\rho^2\bigr) \quad\text{if}\quad \rho^2 = u^2 + v^2 < r_{\text{cut}}^2\,,
  \label{eq:kernel}\\
  \hat{\mathbf{a}} = \textstyle\sum_{i} w_i\,\mathbf{a}_i \,\big/ \sum_{i} w_i\,.
  \label{eq:shepard}
\end{gather}

\begin{figure}[t!]
  \centering
  \includegraphics[width=0.88\linewidth]{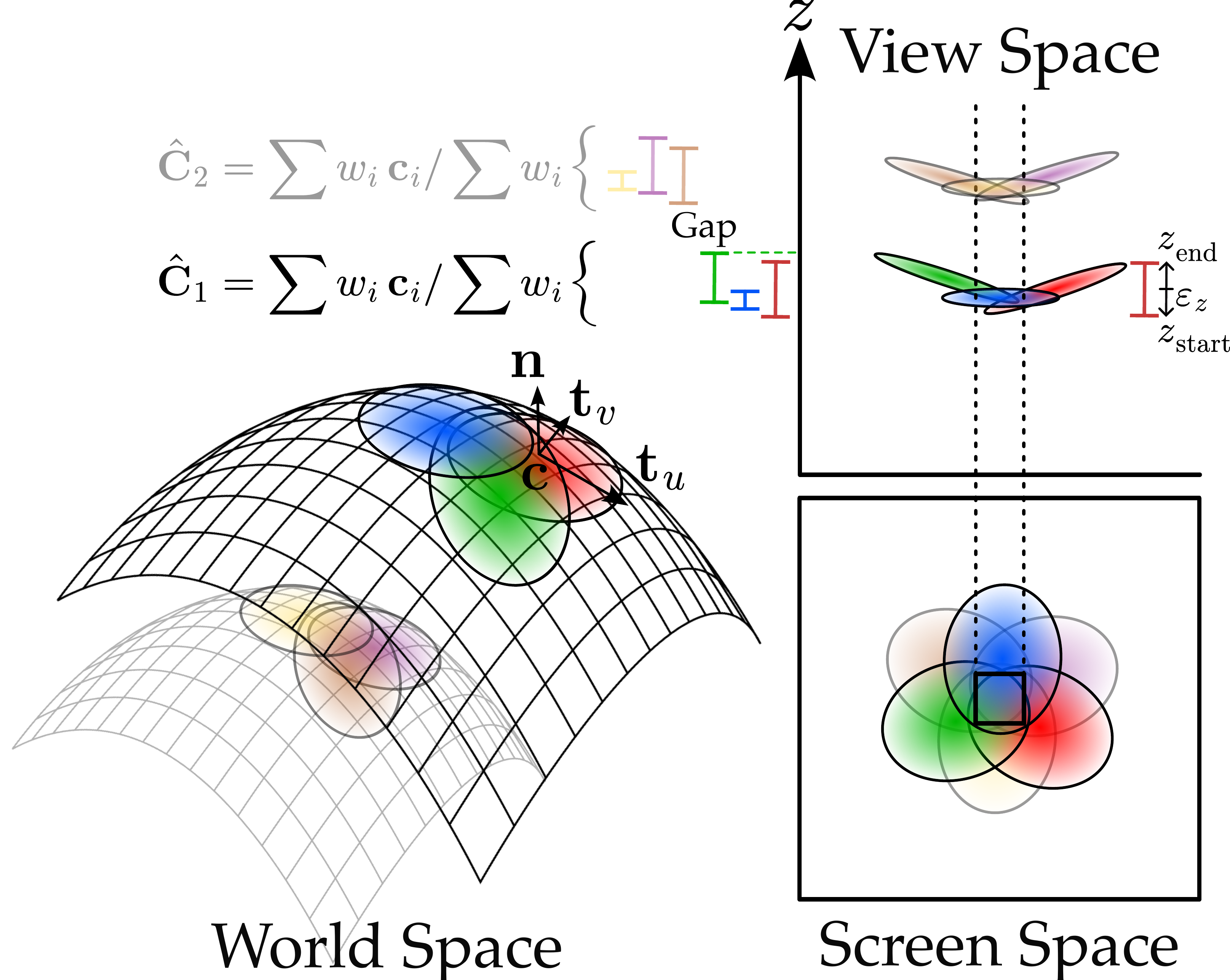}
  \caption{\textbf{Depth-interval grouping.} Surfels sorted by $z_{\text{start}}$ join the current layer while their depth intervals overlap; a gap starts a new one.}
  \label{fig:acc}
\end{figure}

\subsection{Depth-Interval Grouping}
\label{sec:depth_grouping}
\label{sec:interval_grouping}
\Cref{eq:shepard} must average only covering surfels that sample the same surface, or attributes bleed from one surface into another.
Finding them is the \emph{surface separation} problem, which we solve with the kernels themselves: in surface splatting, a surface is not stored anywhere but emerges from the associative overlap of neighboring kernels.
Since a kernel is cut off at $r_{\text{cut}}$, each surfel occupies the depth interval $[z - \varepsilon_z,\, z + \varepsilon_z]$, where $z$ is the camera-space depth of its center and $\varepsilon_z = r_{\text{cut}}\,\|(t'_{u,z},\, t'_{v,z})\|$ the depth extent of its support.
Overlapping intervals merge into one surface layer, and a gap in depth starts a new one: surface separation becomes interval merging over the sorted surfel stream.

To merge intervals exactly in a single pass, we sort the surfels of each tile by interval start $z_{\text{start}} = z - \varepsilon_z$ instead of the center depth used by 3DGS: once a surfel starts beyond the farthest end of the current layer, so do all later ones, and the gap is final without look-ahead.
Each pixel scans its sorted surfels (\cref{fig:acc} and \cref{alg:interval_grouping}), keeping accumulators for the current layer and the running maximum $z^{\max}_{\text{end}}$ of its members' interval ends $z + \varepsilon_z$.
A covering surfel with $z_{\text{start}} > z^{\max}_{\text{end}}$ flushes the layer to the compositor (\cref{sec:compositing}) and opens a new one; any other extends $z^{\max}_{\text{end}}$ and is accumulated.
The same rule separates surfaces along the secondary rays of our Monte Carlo shading (details in \cref{sec:supp_tracing}).

\begin{figure}[t!]
  \centering
  \includegraphics[width=0.86\linewidth]{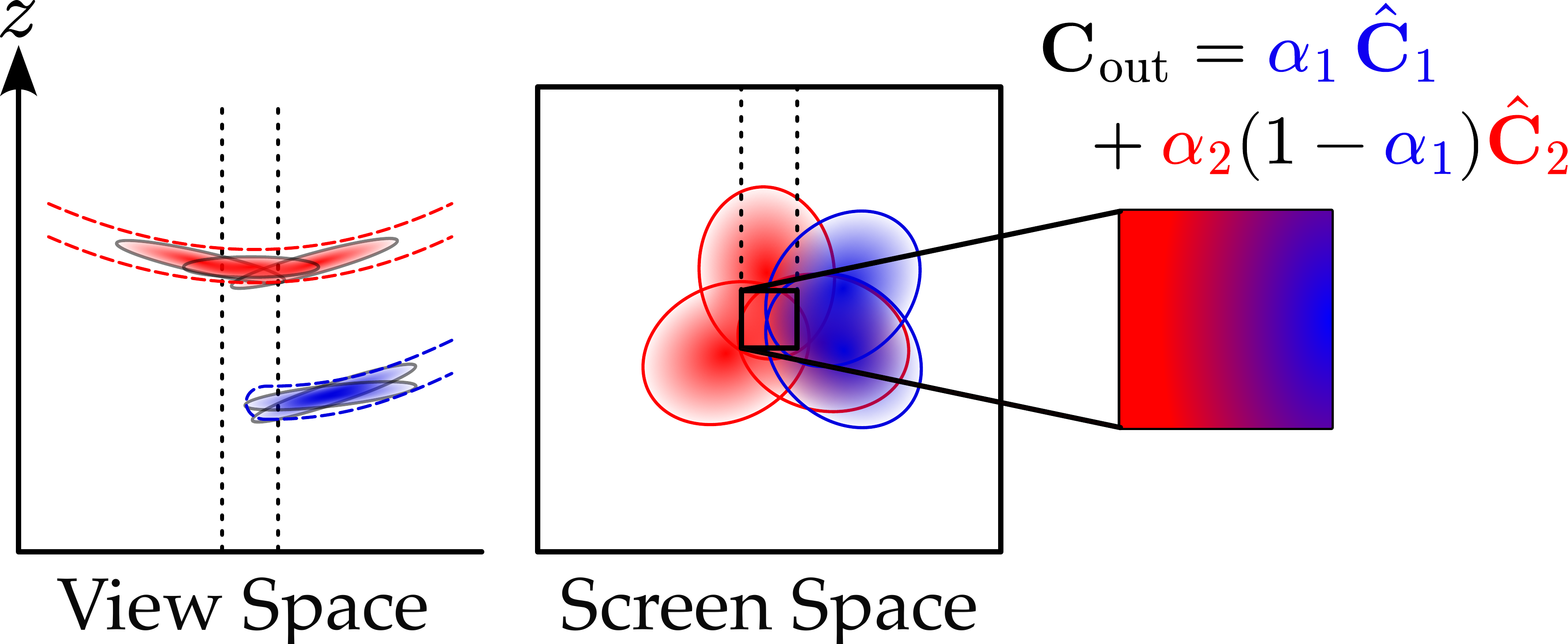}
  \caption{\textbf{Layer compositing.} Layers are composited front to back with their coverage opacities $\alpha_k$.}
  \label{fig:comp}
\end{figure}

\subsection{Coverage-Based Multi-Layer Compositing}
\label{sec:compositing}
\label{sec:coverage}
Grouping yields at each pixel surface layers $k = 1, \ldots, K$ with normalized colors $\hat{\mathbf{C}}_k$ (\cref{eq:shepard}) and accumulated weights $W_k = \sum_{i \in k} w_i$.
Zwicker et al.~\cite{zwicker2001surface} observed that unit-variance Gaussians on a regular grid approximate a partition of unity, so that their sum measures coverage, and used it, with an empirical scale and a clamp, for transparency in an A-buffer~\cite{carpenter1984abuffer}.
We restrict the sum to one surface layer and derive the scale: where surfels approximate a partition of unity, as maintained by adaptive density control (\cref{sec:adc}), a fully covered pixel accumulates about $2\pi$, the kernel's integral (\cref{eq:kernel}), and a pixel covered to a fraction $c$ accumulates $W_k \approx 2\pi c$.
The coverage opacity is this fraction, saturated at full coverage, and layers are composited front to back by the over operator~\cite{porter1984compositing} (\cref{fig:comp}):
\begin{gather}
  \alpha_k = \min\!\bigl(1,\; W_k / 2\pi\bigr)\,,
  \label{eq:alpha}\\
  \mathbf{C}_{\text{out}} = \textstyle\sum_{k=1}^{K} \bar{T}_k\,\alpha_k\,\hat{\mathbf{C}}_k\,,\qquad
  \bar{T}_k = \prod_{j < k}(1 - \alpha_j)\,,
  \label{eq:compositing}
\end{gather}
where the residual transmittance $\bar{T}_k$ is the fraction of the pixel not covered by the layers in front.
At partially covered pixels, along silhouettes and occlusion boundaries, $\alpha_k$ varies continuously with the surfel positions and shapes through the kernel weights, making the image differentiable through visibility for a fixed layer assignment (\cref{sec:limitations}).

\subsection{Object-Space Anti-Aliasing}
\label{sec:mip_filtering}
\label{sec:obj_mip}
Since our kernels are evaluated in object space, so must be their band limiting.
AA-2DGS~\cite{younes2025anti} provides an object-space MIP filter: a screen-space Gaussian pixel prefilter of covariance $\sigma_f\mathbf{I}$, mapped into object space through the Jacobian $\mathbf{J} = \partial(u,v)/\partial(x,y)$ of the ray--surfel intersection, gives the kernel the covariance and amplitude
\begin{equation}
  \boldsymbol{\Sigma}_{\text{mip}} = \mathbf{I} + \sigma_f\,\mathbf{J}\mathbf{J}^\top\,,\qquad
  n_f = \sqrt{|\mathbf{I}|/|\boldsymbol{\Sigma}_{\text{mip}}|}\,.
  \label{eq:mip_cov}
\end{equation}
The kernel thus keeps its integral, on which coverage relies (\cref{eq:alpha}), widens under minification, and reverts to \cref{eq:kernel} under magnification.
\label{sec:center_precomp}%
AA-2DGS evaluates $\mathbf{J}$ at every intersection.
Our \emph{center-precomputed MIP filter} evaluates it once, at the surfel center, since $\mathbf{J}$ varies little over a surfel's small screen footprint, and precomputes $\boldsymbol{\Sigma}^{-1}_{\text{mip}}$ and $n_f$ per surfel; each intersection then evaluates, instead of \cref{eq:kernel},
\begin{equation}
  w = n_f \exp\!\bigl(-\tfrac{1}{2}\,\rho^2_{\text{mip}}\bigr)\,,\qquad
  \rho^2_{\text{mip}} = \mathbf{u}^\top \boldsymbol{\Sigma}^{-1}_{\text{mip}}\,\mathbf{u}\,,
  \label{eq:mip_weight}
\end{equation}
with $\mathbf{u} = (u, v)^\top$ and the cutoff applied to $\rho^2_{\text{mip}}$.
This lowers the per-intersection cost without visible loss (\cref{sec:ablation_mip}).

\subsection{Forward Shading}
\label{sec:forward_shading}

Multi-layer compositing finalizes each surface layer on the fly, so deferred shading would need a G-buffer per layer.
We instead shade each surfel \emph{before} rasterization into a linear HDR color $\mathbf{c}_i$, averaged and composited like any other attribute; the rasterizer is agnostic to the shading function, which differs between our two optimization stages (\cref{sec:optimization}), as they solve different problems.
In the geometry stage, $\mathbf{c}_i$ is the surfel's SH radiance, as in the pre-training of IRGS~\cite{gu2024IRGS}; unconstrained by physics, it absorbs view-dependent appearance that would otherwise distort the geometry.
We also render a physically-based image with split-sum image-based lighting~\cite{karis2013real,Munkberg_2022_CVPR} against a learnable cubemap; this makes the geometry also explain a microfacet BRDF (details in \cref{sec:forward_shading_practice}).

\paragraph{Monte Carlo shading.}
\label{sec:mc_shading}
In the inverse-rendering stage, on frozen geometry, $\mathbf{c}_i$ is a Monte Carlo estimate of the rendering equation from $S$ directions $\boldsymbol{\omega}_s$ of density $p$,
\begin{gather}
  \mathbf{c}_i = \tfrac{1}{S}\textstyle\sum_{s}
    (f_d + f_s(\boldsymbol{\omega}_s))\,L(\boldsymbol{\omega}_s)\,(\mathbf{n}\cdot\boldsymbol{\omega}_s)^{+} / p(\boldsymbol{\omega}_s),
  \label{eq:mc_shading}\\
  L = V L_{\text{env}} + L_{\text{ind}}\,,
  \label{eq:incident}
\end{gather}
with a Lambertian lobe $f_d = (1-m)\,\boldsymbol{\rho}/\pi$, a GGX lobe $f_s$~\cite{walter2007microfacet} of base reflectivity $\mathbf{F}_0 = 0.04\,(1-m) + m\,\boldsymbol{\rho}$, and incident radiance $L$: a learnable equirectangular environment map $L_{\text{env}}$, attenuated by the visibility $V$, plus the indirect radiance $L_{\text{ind}}$.
\label{sec:sampling}%
We use $S = 64$ directions of a Fibonacci lattice, randomly rotated once per surfel since the trace is cached (\cref{sec:supp_caching}), each carrying the weight $2\pi$ of uniform hemisphere sampling; at evaluation, part of them importance-sample the environment map, weighted by the balance heuristic~\cite{veach1995optimally}.
A uniform set suits the diffuse lobe better than a narrow specular one, which biases the estimated roughness upward (details in \cref{sec:supp_mc}).

\paragraph{Secondary rays.}
\label{sec:secondary_rays}
$V$ and $L_{\text{ind}}$ come from secondary rays resolved with the same surface model as camera rays: each ray is intersected in closed form with candidates from a bounding volume hierarchy (BVH) over the support disks, each hit spans the interval that its support covers along the ray (computed like $\varepsilon_z$), and overlapping intervals merge into surface layers by the same rule as in \cref{sec:depth_grouping}.
The trace skips the layer of the originating surface and accumulates the next one, up to the gap that closes it.
With this layer's Shepard-normalized radiance $\hat{\mathbf{C}}$ and coverage $\alpha$ (\cref{eq:alpha}),
\begin{equation}
  V = 1 - \alpha\,, \qquad L_{\text{ind}} = \alpha\,\hat{\mathbf{C}}\,.
  \label{eq:trace}
\end{equation}
$\hat{\mathbf{C}}$ is read from the SH radiance of the hit surfels, so the radiance field fitted in the geometry stage caches outgoing radiance in directions no camera observed.
Only this first layer contributes indirect light; \cref{sec:supp_tracing} details the BVH and an order-independent interval merge without per-ray sorting.

\subsection{Optimization}
\label{sec:optimization}
\begin{table*}[t]
\centering
\scriptsize
\setlength{\tabcolsep}{4pt}
\renewcommand{\arraystretch}{1.0}
\caption{\textbf{Inverse rendering on Stanford-ORB.} \textdagger: surfels sampled on the ground-truth scans. Shape: Chamfer distance $\times 10^{-3}$.}
\label{tab:benchmark}
\begin{tabular}{@{}cl cc cccc cccc@{}}
\toprule
 & \multirow{2}{*}{}
 & \multicolumn{2}{c}{Geometry}
 & \multicolumn{4}{c}{Novel Scene Relighting}
 & \multicolumn{4}{c}{Novel View Synthesis}
 \\
 \cmidrule(l){3-4} \cmidrule(l){5-8} \cmidrule(l){9-12}
& & Normal$\downarrow$ & Shape$\downarrow$ & PSNR-H$\uparrow$ & PSNR-L$\uparrow$ & SSIM$\uparrow$ & LPIPS$\downarrow$ & PSNR-H$\uparrow$ & PSNR-L$\uparrow$ & SSIM$\uparrow$ & LPIPS$\downarrow$\\
\midrule
\multirow{5}{*}{\rotatebox[origin=c]{90}{\scriptsize GS-based}}
& GS-IR~\cite{liang2024gs}              & -- & -- & -- & 25.98 & 0.897 & 0.092 & -- & 34.80 & 0.960 & 0.047\\
& R3DG~\cite{gao2023relightable}        & -- & -- & -- & 28.52 & 0.931 & 0.069 & -- & 37.15 & 0.981 & 0.024\\
& GaussianShader~\cite{jiang2024gaussianshader} & -- & -- & -- & 26.86 & 0.930 & 0.063 & -- & 37.13 & \third{0.982} & \third{0.023}\\
& RTR-GS~\cite{zhou2025rtr}             & -- & -- & -- & 30.10 & 0.944 & 0.053 & -- & \second{39.17} & \second{0.985} & \second{0.021}\\
& IRGS~\cite{gu2024IRGS}                & \second{0.05} & \second{0.51} & \second{25.20} & \second{32.49} & \second{0.974} & \second{0.032} & \second{28.52} & \third{37.17} & \third{0.982} & 0.025\\
\midrule
\multirow{3}{*}{\rotatebox[origin=c]{90}{\scriptsize Implicit}}
& PhySG~\cite{zhang2021physg}         & 0.17 & 9.28  & 21.81 & 28.11 & 0.960 & 0.055 & 24.24 & 32.15 & 0.974 & 0.047\\
& NeRFactor~\cite{zhang2021nerfactor} & 0.29 & 9.53  & 23.54 & 30.38 & 0.969 & 0.048 & 26.06 & 33.47 & 0.973 & 0.046\\
& InvRender~\cite{zhang2022invrender}        & \third{0.06} & \first{0.44} & 23.76 & 30.83 & 0.970 & 0.046 & 25.91 & 34.01 & 0.977 & 0.042\\
\midrule
\multirow{2}{*}{\rotatebox[origin=c]{90}{\scriptsize Mesh}}
& NVDiffRec~\cite{Munkberg_2022_CVPR}   & \third{0.06} & \third{0.62} & 22.91 & 29.72 & 0.963 & 0.039 & 21.94 & 28.44 & 0.969 & 0.030\\
& NVDiffRecMC~\cite{hasselgren2022shape} & \first{0.04} & \second{0.51} & \third{24.43} & \third{31.60} & \third{0.972} & \third{0.036} & \third{28.03} & 36.40 & \third{0.982} & 0.028\\
\midrule
\multirow{2}{*}{}
& 3DSS\textsuperscript{\textdagger} & 0.00 & 0.01 & 25.73 & 33.13 & 0.978 & 0.026 & 29.05 & 37.95 & 0.988 & 0.016\\
& 3DSS (Ours)       & \third{0.06} & \first{0.44} & \first{25.50} & \first{32.79} & \first{0.978} & \first{0.023} & \first{30.84} & \first{40.07} & \first{0.991} & \first{0.011}\\
\bottomrule
\end{tabular}
\end{table*}

\begin{table*}[t]
\centering
\footnotesize
\setlength{\tabcolsep}{4pt}
\renewcommand{\arraystretch}{1.0}
\caption{\textbf{Inverse rendering on TensoIR Synthetic}, averaged over the four scenes (Armadillo, Ficus, Hotdog, Lego).}
\label{tab:tensoir}
\begin{tabular}{@{}cl ccc ccc ccc c@{}}
\toprule
 & \multirow{2}{*}{}
 & \multicolumn{3}{c}{Novel View Synthesis}
 & \multicolumn{3}{c}{Albedo}
 & \multicolumn{3}{c}{Relighting}
 & \multirow{2}{*}{Runtime}
 \\
 \cmidrule(l){3-5} \cmidrule(l){6-8} \cmidrule(l){9-11}
& & PSNR$\uparrow$ & SSIM$\uparrow$ & LPIPS$\downarrow$ & PSNR$\uparrow$ & SSIM$\uparrow$ & LPIPS$\downarrow$ & PSNR$\uparrow$ & SSIM$\uparrow$ & LPIPS$\downarrow$ & \\
\midrule
\multirow{3}{*}{\rotatebox[origin=c]{90}{\scriptsize Implicit}}
& NeRFactor~\cite{zhang2021nerfactor} & 24.679 & 0.922 & 0.120 & 25.125 & 0.940 & 0.109 & 23.383 & 0.908 & 0.131 & $>$100 hrs\\
& InvRender~\cite{zhang2022invrender} & 27.367 & 0.934 & 0.089 & 27.341 & 0.933 & 0.100 & 23.973 & 0.901 & 0.101 & 15 hrs\\
& TensoIR~\cite{jin2023tensoir} & 35.088 & \second{0.976} & \third{0.040} & 29.275 & \third{0.950} & 0.085 & 28.580 & \second{0.944} & \third{0.081} & 5 hrs\\
\midrule
\multirow{2}{*}{\rotatebox[origin=c]{90}{\scriptsize Mesh}}
& NVDiffRec~\cite{Munkberg_2022_CVPR} & 30.696 & 0.962 & 0.052 & 29.174 & 0.908 & 0.115 & 19.880 & 0.879 & 0.104 & $<$1 hr\\
& NVDiffRecMC~\cite{hasselgren2022shape} & 33.132 & 0.953 & 0.071 & \second{32.563} & \first{0.955} & \first{0.071} & \third{29.312} & 0.928 & 0.090 & $<$2 hrs\\
\midrule
\multirow{2}{*}{\rotatebox[origin=c]{90}{\scriptsize\shortstack{GS-\\based}}}
& GS-IR~\cite{liang2024gs} & \third{35.333} & \third{0.974} & \second{0.039} & 30.286 & 0.941 & 0.084 & 24.374 & 0.885 & 0.096 & $<$1 hr\\
& IRGS~\cite{gu2024IRGS} & \second{35.518} & 0.964 & 0.049 & \first{33.401} & \second{0.954} & \second{0.076} & \second{30.631} & \third{0.935} & \second{0.076} & $<$1 hr\\
\midrule
& 3DSS (Ours) & \first{36.953} & \first{0.977} & \first{0.029} & \third{31.440} & 0.943 & \third{0.083} & \first{31.681} & \first{0.951} & \first{0.055} & $<$1 hr\\
\bottomrule
\end{tabular}
\end{table*}

Following IRGS~\cite{gu2024IRGS}, we optimize with Adam~\cite{kingma2014adam} in two stages that differ in which parameters are free.
The geometry stage optimizes the geometry, SH radiance, materials, and cubemap under adaptive density control.
The inverse-rendering stage freezes the geometry and re-initializes the materials and the environment, which so far served only as a proxy for geometry; the SH radiance keeps training, because secondary rays read it.

\paragraph{Initialization.}
\label{sec:init}
Surfels are unprojected from the depth maps that Depth Anything~3~\cite{lin2026depth} predicts for the training views, oriented by the depth gradients, and given isotropic scales from the nearest-neighbor distance so that kernels overlap; a sphere of surfels is the prior-free alternative (\cref{sec:ablation_components}).

\paragraph{Adaptive density control.}
\label{sec:adc}
As in 3DGS~\cite{kerbl2023gaussian}, densification is guided by the gradient of the loss with respect to the projected surfel centers, obtained analytically through the T-matrix.
A surfel with a large gradient is split into two coplanar children along its major axis if its larger scale exceeds the mean distance to its neighbors; unlike the global scene-extent threshold of 3DGS, this local criterion keeps the kernel supports overlapping, as the coverage model (\cref{sec:coverage}) requires.
Surfels with no gradient, no neighbor within their support, or outside the masks are pruned (details in \cref{sec:density_control}).

\paragraph{Losses.}
\label{sec:loss_function}
The photometric losses $\mathcal{L}^{\text{sh}}$ and $\mathcal{L}^{\text{pbr}}$ of the two images are $(1-\lambda)\,\ell_1 + \lambda\,\text{D-SSIM}$ on tone-mapped images, log- and sRGB-transformed for HDR targets~\cite{Munkberg_2022_CVPR} so that bright pixels do not dominate.
With masks, the silhouette loss $\mathcal{L}_{\text{sil}}$ is the mean $\ell_1$ distance between the total coverage $1 - \bar{T}_{K+1}$ and the fractional ground-truth alpha.
Multi-layer compositing lets the optimizer spread a surface over spurious depth layers, which a depth consolidation loss, inspired by the distortion losses of Mip-NeRF~360~\cite{barron2022mip} and 2DGS~\cite{huang20242d}, counters per pixel:
\begin{equation}
  \mathcal{L}_{\text{cons}} = \textstyle\sum_{k<l} \bar{T}_k\alpha_k\,\bar{T}_l\alpha_l\,|z_k - z_l| + \sum_{k} (\bar{T}_k\alpha_k)^2\,\sigma_k^2\,,
  \label{eq:cons_loss}
\end{equation}
where layer~$k$ has compositing weight $\bar{T}_k\alpha_k$, mean depth $z_k$, and depth variance $\sigma_k^2$: the first term gathers the weight of a pixel into one layer, the second thins each layer.

\label{sec:normal_reg}%
A normal consistency loss $\mathcal{L}_{\text{nc}}$ aligns the normal of each surfel with those of its nearest neighbors; being local, it is blind to a smooth but globally misoriented patch, which a depth-normal consistency loss $\mathcal{L}_{\text{nd}}$ catches by comparing, in screen space, the rendered normals with those of the rendered depth.
A white-light prior, total variation on the environment map, and edge-aware smoothness of albedo and roughness regularize the material--light split, as in R3DG~\cite{gao2023relightable} and IRGS~\cite{gu2024IRGS} (details in \cref{sec:supp_losses}).

\section{Experiments}
\label{sec:experiments}

\begin{figure*}[t]
  \centering
  \setlength{\tabcolsep}{0.5pt}
  \renewcommand{\arraystretch}{0.6}
  \newcommand{\RLw}{0.12\textwidth}
  \newcommand{\RLdir}{Figures/Comparisons/TensoIR/relight}
  \newcommand{\RLgroup}[1]{%
    \figimg[width=\RLw]{\RLdir/#1/nvdiffrecmc.jpg} &
    \figimg[width=\RLw]{\RLdir/#1/irgs.jpg} &
    \figimg[width=\RLw]{\RLdir/#1/ours.jpg} &
    \figimg[width=\RLw]{\RLdir/#1/gt.jpg}}
  \newcommand{\RLrow}[5]{%
    #5 & \RLgroup{#1} & \RLgroup{#2} \\[0.5pt]
    & \multicolumn{4}{c}{\scriptsize\emph{#3}} & \multicolumn{4}{c}{\scriptsize\emph{#4}}}
  \newcommand{\RLlabel}[1]{\smash{\rotatebox{90}{\makebox[\dimexpr\RLw*2+9pt\relax][c]{\small #1}}}}
  \newcommand{\RLhead}{{\scriptsize NVDiffRecMC} & {\scriptsize IRGS} & {\scriptsize \textbf{3DSS (Ours)}} & {\scriptsize GT}}
  \begin{tabular}{@{}c cccc@{\hskip 5pt}cccc@{}}
  & \RLhead & \RLhead \\[1pt]
  \RLrow{armadillo_test_080_forest}{armadillo_test_074_city}{forest}{city}{} \\[1pt]
  \RLrow{armadillo_test_059_fireplace}{armadillo_test_124_bridge}{fireplace}{bridge}{\RLlabel{Armadillo}} \\[4pt]
  \RLrow{lego_test_191_forest}{lego_test_196_city}{forest}{city}{} \\[1pt]
  \RLrow{lego_test_167_fireplace}{lego_test_001_bridge}{fireplace}{bridge}{\RLlabel{Lego}} \\
  \end{tabular}
  \caption{\textbf{Relighting on TensoIR Synthetic.} All methods are shown side by side under each held-out environment map.}
  \label{fig:tensoir_relight}
\end{figure*}

\paragraph{Datasets and metrics.}
We evaluate inverse rendering from multi-view images under unknown illumination.
Stanford-ORB~\cite{kuang2023stanfordorb} has 14 real objects captured in seven scenes, with laser scans, pseudo-albedo, calibrated HDR images, and measured lighting; each object is relit under its other scenes' lighting.
We score normals by cosine distance, shape by Chamfer distance, and relighting and novel view synthesis (NVS) by HDR and LDR PSNR (PSNR-H/L), SSIM, and LPIPS; since our surfels form an oriented point cloud, we mesh them with NKSR (Neural Kernel Surface Reconstruction)~\cite{huang2023nksr}, a feed-forward 3D implicit surface reconstruction method, rather than by depth fusion.
TensoIR Synthetic~\cite{jin2023tensoir} has four Blender objects with thin structures, complex materials, and self-occlusion (shadows, interreflections); each has 100 training and 200 test views ($800{\times}800$), ground-truth albedo and normals, and five held-out environment maps.
We score NVS, per-channel scale-aligned albedo~\cite{zhang2021nerfactor}, and relighting averaged over the five maps by PSNR, SSIM, and LPIPS, and normals by mean angular error (MAE) in degrees.

\paragraph{Baselines.}
We compare with implicit (PhySG~\cite{zhang2021physg}, NeRFactor~\cite{zhang2021nerfactor}, InvRender~\cite{zhang2022invrender}, TensoIR~\cite{jin2023tensoir}), mesh (NVDiffRec~\cite{Munkberg_2022_CVPR}, NVDiffRecMC~\cite{hasselgren2022shape}), and Gaussian splatting methods (GS-IR~\cite{liang2024gs}, R3DG~\cite{gao2023relightable}, GaussianShader~\cite{jiang2024gaussianshader}, RTR-GS~\cite{zhou2025rtr}, IRGS~\cite{gu2024IRGS}), mainly the 2DGS-based IRGS and NVDiffRecMC, which we retrain.
The fixed-geometry configuration 3DSS\textsuperscript{\textdagger} samples surfels uniformly on the Stanford-ORB scans and optimizes only shading and lighting, to isolate shading fidelity and show compatibility with mesh assets.
Details are in \cref{sec:supp_setup}, and additional qualitative comparisons in \cref{sec:supp_qualitative}.

\subsection{Inverse Rendering Comparisons}
\label{sec:main_results}

\begin{figure}[t]
  \centering
  \setlength{\tabcolsep}{0.6pt}
  \renewcommand{\arraystretch}{0.6}
  \newlength{\FICw}%
  \setlength{\FICw}{\dimexpr(\linewidth-14pt)*94/400\relax}%
  \newcommand{\FIC}[1]{\figimg[width=\FICw]{Figures/Comparisons/TensoIR/ficus_test_053/#1.png}}%
  \newcommand{\FICrow}[2]{%
    \rotatebox{90}{\makebox[\FICw][c]{\scriptsize #1}} &
    \FIC{nvdiffrecmc_#2} & \FIC{irgs_#2} & \FIC{ours_#2} & \FIC{gt_#2}}%
  \begin{tabular}{@{}c cccc@{}}
  & {\scriptsize NVDiffRecMC} & {\scriptsize IRGS} & {\scriptsize \textbf{3DSS (Ours)}} & {\scriptsize GT} \\[1pt]
  \FICrow{Normal}{normal_box} \\[2pt]
  \FICrow{Zoom}{zoomA} \\
  \FICrow{Zoom}{zoomB} \\[2pt]
  \FICrow{Albedo}{albedo_crop} \\
  \FICrow{Rendering}{rgb_crop} \\
  \end{tabular}
  \caption{\textbf{Normals, albedo, and NVS on TensoIR Synthetic.} Boxed regions are magnified below the normals: NVDiffRecMC merges leaflets into blobs and drops thin stems, which 3DSS preserves. Each albedo channel is scaled by a global factor, following NeRFactor~\cite{zhang2021nerfactor}.}
  \label{fig:tensoir_qualitative}
\end{figure}

\begin{figure}[t]
  \centering
  \setlength{\tabcolsep}{0.5pt}
  \renewcommand{\arraystretch}{0.6}
  \newlength{\GEOw}%
  \setlength{\GEOw}{\dimexpr(\linewidth-14pt)/4\relax}%
  \newcommand{\GEOrow}[4]{%
    \rotatebox{90}{\makebox[\dimexpr\GEOw*#4/1000\relax][c]{\small #3}} &
    \figimg[width=\GEOw,trim=#2,clip]{Figures/Comparisons/Geometry/#1/nvdiffrecmc_mesh.png} &
    \figimg[width=\GEOw,trim=#2,clip]{Figures/Comparisons/Geometry/#1/irgs_tsdf.png} &
    \figimg[width=\GEOw,trim=#2,clip]{Figures/Comparisons/Geometry/#1/3dss_nksr.png} &
    \figimg[width=\GEOw,trim=#2,clip]{Figures/Comparisons/Geometry/#1/gt_mesh.png}}%
  \begin{tabular}{@{}ccccc@{}}
    & {\footnotesize NVDiffRecMC} & {\footnotesize IRGS (TSDF)} &
      {\footnotesize 3DSS (NKSR)} & {\footnotesize GT} \\[1pt]
    \GEOrow{gnome_scene003}{20 11 20 11}{Gnome}{1050} \\
    \GEOrow{teapot_scene006}{18 4 17 5}{Teapot}{1137} \\
  \end{tabular}
  \caption{\textbf{Meshes on Stanford-ORB} against ground-truth scans.}
  \label{fig:geometry}
\end{figure}

\begin{figure}[t]
  \centering
  \setlength{\tabcolsep}{1pt}
    \begin{tabular}{@{}ccc@{}}
      \includegraphics[width=0.32\linewidth]{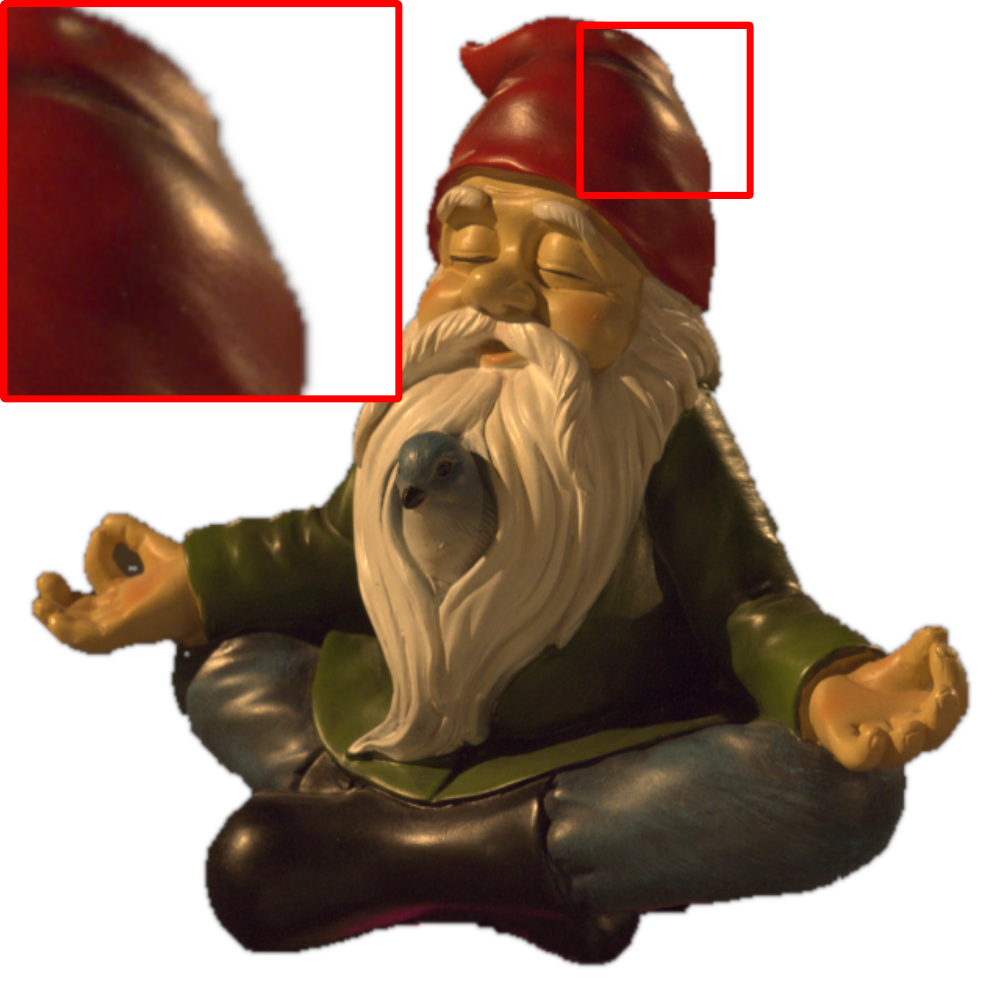} &
      \includegraphics[width=0.32\linewidth]{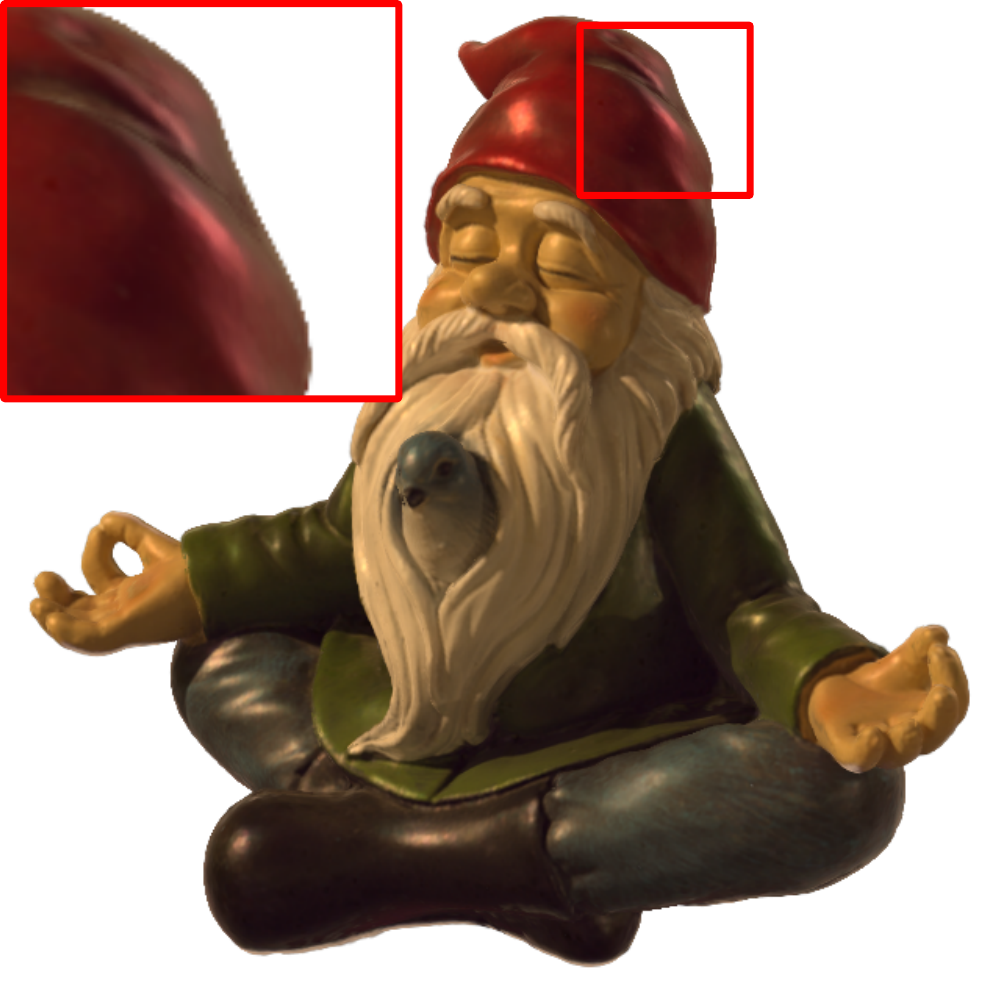} &
      \includegraphics[width=0.32\linewidth]{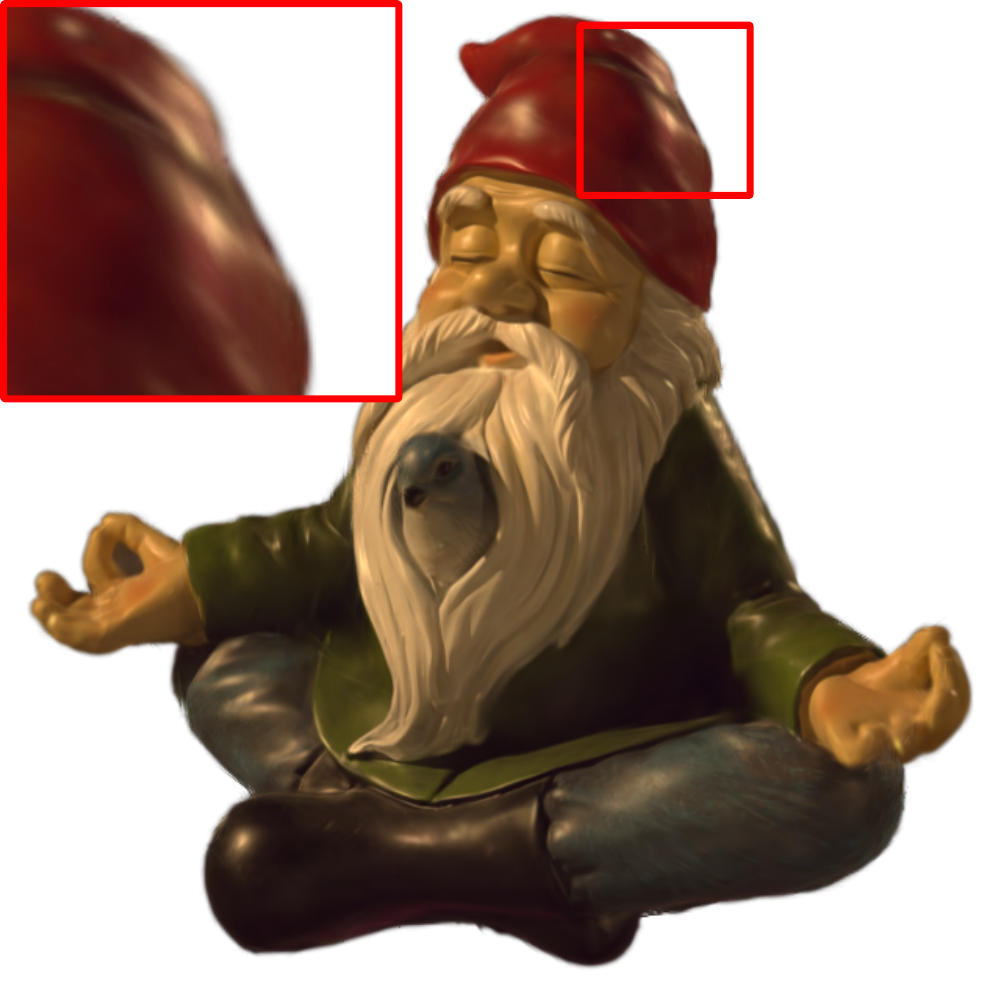} \\[1pt]
      {\small (a) Ground Truth} &
      {\small (b) 3DSS\textsuperscript{\textdagger} (1M)} &
      {\small (c) 3DSS (100K)} \\
    \end{tabular}
  \caption{\textbf{Fixed geometry.} \textbf{(b)} has exact geometry and $10{\times}$ more surfels, yet its uniform sampling blurs specular reflections and material detail that the adaptively densified \textbf{(c)} recovers.}
  \label{fig:fixed_vs_optimized}
\end{figure}

\begin{figure}[t]
  \centering
  \setlength{\tabcolsep}{0.4pt}
  \renewcommand{\arraystretch}{0.5}
  \newlength{\orbh}%
  \setlength{\orbh}{\dimexpr(\linewidth-16pt)*950/3560\relax}%
  \newcommand{\ORBt}[3]{\figimg[height=\orbh,trim=#3,clip]{Figures/Comparisons/#1_#2.png}}%
  \newcommand{\ORBrow}[2]{%
    \rotatebox{90}{\makebox[\orbh][c]{\scriptsize #2}} &
    \ORBt{ORB/baking/nvs}{#1}{6 3 5 3} &
    \ORBt{ORB/baking/albedo}{#1}{6 3 5 3} &
    \ORBt{ORB/baking/relight}{#1}{7 8 8 8} &
    \ORBt{ORB/cactus/nvs}{#1}{43 18 44 18} &
    \ORBt{ORB/cactus/albedo}{#1}{43 18 44 18} &
    \ORBt{ORB/cactus/relight}{#1}{18 29 18 29}}%
  \begin{tabular}{@{}c ccc@{\hskip 2.5pt}ccc@{}}
  & \multicolumn{3}{c}{\small Baking} & \multicolumn{3}{c}{\small Cactus} \\[-1pt]
  \cmidrule(lr){2-4} \cmidrule(lr){5-7}
  & {\scriptsize NVS} & {\scriptsize Albedo} & {\scriptsize Relight}
  & {\scriptsize NVS} & {\scriptsize Albedo} & {\scriptsize Relight} \\
  \ORBrow{nvdiffrecmc}{NVDiffRecMC} \\
  \ORBrow{irgs}{IRGS} \\
  \ORBrow{3dss}{\textbf{3DSS (Ours)}} \\
  \ORBrow{gt}{GT} \\
  \end{tabular}
  \caption{\textbf{Qualitative comparison on Stanford-ORB}; relighting shows another view. Albedo and relighting are scale-aligned.}
  \label{fig:main_comparison}
\end{figure}

\paragraph{Relighting.}
\label{sec:relighting_results}
3DSS relights best on both benchmarks (\cref{tab:benchmark,tab:tensoir}).
Its lead over IRGS is small in PSNR and larger in LPIPS on Stanford-ORB, and grows on the self-shadowing TensoIR objects, where GS-IR, with baked occlusion, falls well behind.
In \cref{fig:tensoir_relight}, NVDiffRecMC and IRGS make the Armadillo glossy and plastic, whereas 3DSS keeps its diffuse look and shell relief; on the Lego, 3DSS resolves the track links that NVDiffRecMC merges and IRGS blurs.
On real data (\cref{fig:main_comparison}), IRGS adds wrong highlights on Cactus; only 3DSS keeps the Baking label legible.

\paragraph{Novel view synthesis.}
\label{sec:nvs_results}
3DSS leads every NVS metric of both benchmarks.
Mesh methods trail the strongest Gaussian methods, even NVDiffRecMC, which relights well; we attribute this to the representation, not the shading: a fixed-resolution tetrahedral grid caps a mesh's triangle budget, whereas points are densified where the image demands it, an adaptivity that 3DSS keeps while rendering an opaque surface.
In \cref{fig:main_comparison}, NVDiffRecMC flattens the Cactus vase and IRGS blurs highlights and geometry, whereas 3DSS reproduces the fine print of the Baking label and the Ficus pot ridges that IRGS smooths away (\cref{fig:tensoir_qualitative}).

\paragraph{Materials.}
\label{sec:material_results}
3DSS ranks third in albedo PSNR and LPIPS (\cref{tab:tensoir}), behind IRGS and NVDiffRecMC; the latter's albedo is oversmoothed but free of shading, which the scale-aligned metrics reward, whereas IRGS and 3DSS keep some residual shading (\cref{fig:tensoir_qualitative}).
All three bake the Ficus pot ridges, which IRGS and 3DSS miss in the normals, into the albedo, ours most visibly, and NVDiffRecMC tints the soil and stems green.
On the real objects (\cref{fig:main_comparison}), 3DSS preserves the print of the Baking label, which NVDiffRecMC largely loses and IRGS blurs.

\paragraph{Geometry.}
\label{sec:geometry_results}
On Stanford-ORB, 3DSS ties InvRender for the lowest Chamfer distance (\cref{tab:benchmark}).
In the TensoIR normals (\cref{fig:tensoir_qualitative}), on the Ficus, the NVDiffRecMC mesh misses branches finer than its grid and merges leaves into blobs, whereas 3DSS, though a surface representation, comes close to the volumetric IRGS, since surfels are bound to neither a grid nor a connectivity.
In \cref{fig:geometry}, IRGS's TSDF meshes show fusion artifacts at the Teapot base, while ours recover the Gnome's beard strands and the smooth Teapot.

\paragraph{Fixed geometry.}
\label{sec:frozen_gt_results}
The near-zero geometry error of 3DSS\textsuperscript{\textdagger} shows that surfel sampling and rendering add negligible distortion.
It beats NVDiffRecMC and IRGS on every NVS metric and RTR-GS on all but PSNR-L.
Its relighting PSNR is the highest, above the full 3DSS at equal SSIM: with correct geometry, our shading recovers better materials, so part of the full model's relighting error stems from its geometry, not the renderer.
In NVS the order reverses, which we attribute to the limitations of the shading model and to the uniform surfel sampling (\cref{fig:fixed_vs_optimized}).

\subsection{Ablation Studies}
\label{sec:ablations}

\begin{table}[t]
\centering
\footnotesize
\setlength{\tabcolsep}{5pt}
\renewcommand{\arraystretch}{1.0}
\caption{\textbf{Ablation on TensoIR Synthetic}: one change per row; relighting over five maps, albedo scaled per channel.}
\label{tab:tensoir_ablation}
\begin{tabular}{@{}lcccc@{}}
\toprule
 & Relighting & NVS & Albedo & Normal \\
 & PSNR$\uparrow$ & PSNR$\uparrow$ & PSNR$\uparrow$ & MAE ($^\circ$)$\downarrow$ \\
\midrule
w/o $\mathcal{L}_{\text{cons}}$ & \second{31.67} & \third{36.78} & \third{31.28} & \second{4.72} \\
w/o $\mathcal{L}_{\text{nc}}$ & 30.70 & \second{36.87} & 30.53 & \third{4.75} \\
w/o $\mathcal{L}_{\text{sil}}$ & \third{31.49} & 36.38 & \second{31.30} & 4.76 \\
w/o splitting & 30.52 & 33.16 & 30.81 & 5.12 \\
Sphere init.\ (200K) & 30.79 & 35.03 & 30.99 & 5.06 \\
\midrule
Full 3DSS & \first{31.68} & \first{36.95} & \first{31.44} & \first{4.60} \\
\bottomrule
\end{tabular}
\end{table}

\begin{figure}[t]
  \centering
  \begin{subfigure}[b]{0.21\linewidth}
    \centering
    \includegraphics[height=74pt, trim=15 6 9 12, clip]{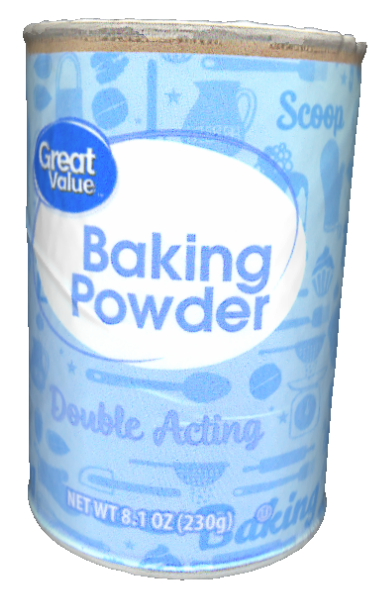}
    \caption{Reference}
    \label{fig:ddist_ref}
  \end{subfigure}\hfill
  \begin{subfigure}[b]{0.37\linewidth}
    \centering
    \includegraphics[height=74pt, trim=15 1 11 2, clip]{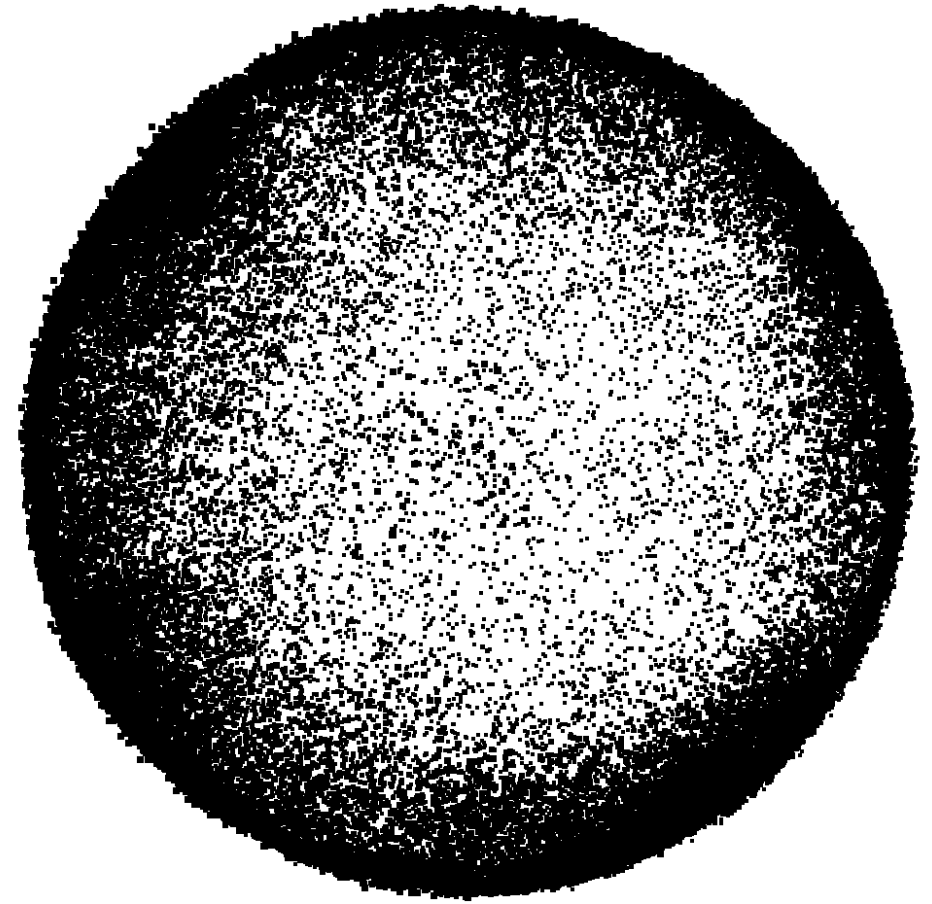}
    \caption{w/o $\mathcal{L}_{\text{cons}}$}
    \label{fig:ddist_without}
  \end{subfigure}\hfill
  \begin{subfigure}[b]{0.37\linewidth}
    \centering
    \includegraphics[height=74pt, trim=15 1 11 2, clip]{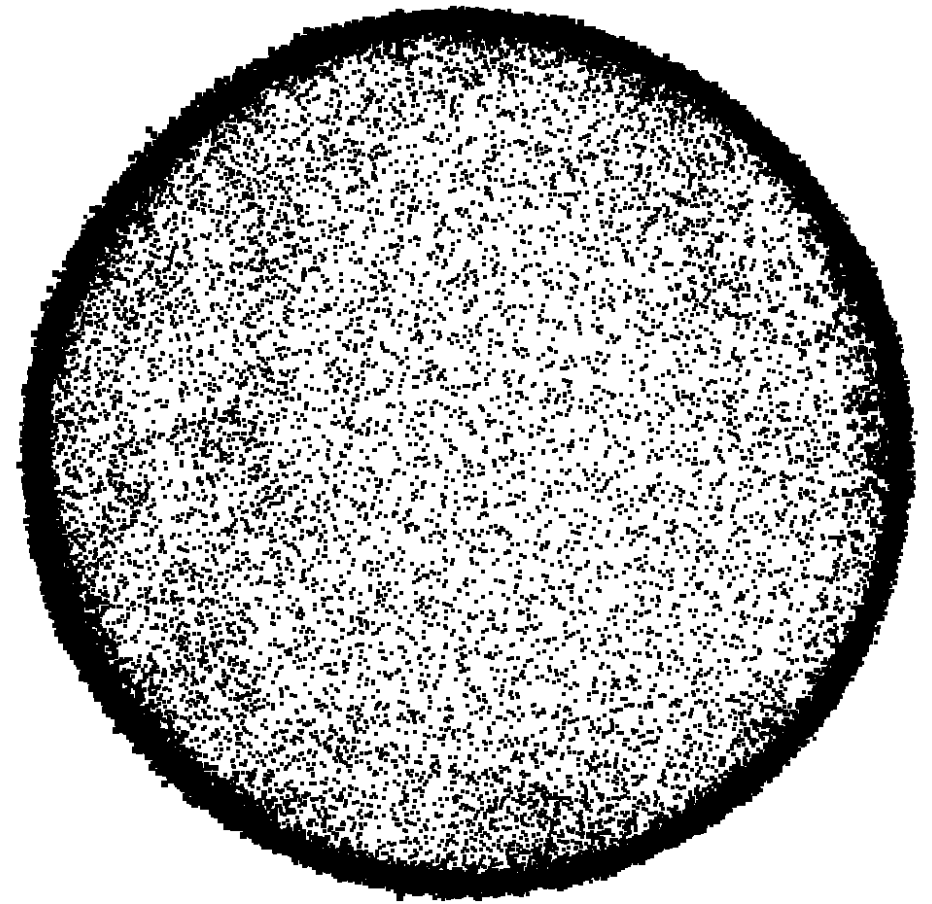}
    \caption{w/ $\mathcal{L}_{\text{cons}}$ (Ours)}
    \label{fig:ddist_with}
  \end{subfigure}
  \caption{\textbf{Depth consolidation.} Top-down cross-sections of the surfels of the Stanford-ORB Baking can~\subref{fig:ddist_ref}: a thick, noisy shell without $\mathcal{L}_{\text{cons}}$~\subref{fig:ddist_without}, a single thin layer with it~\subref{fig:ddist_with}.}
  \label{fig:ddist_ablation}
\end{figure}

\paragraph{Optimization components.}
\label{sec:ablation_components}
Each row of \cref{tab:tensoir_ablation} changes one component at a comparable per-scene surfel budget, so the differences reflect the distribution of the surfels, not their number (details in \cref{sec:supp_ablation}).
Disabling splitting hurts most, above all novel view synthesis: coarse initial surfels cannot represent high-frequency appearance or thin structures.
Each regularizer improves every metric.
Without $\mathcal{L}_{\text{nc}}$, misoriented surfels fit the training views with compensating albedo or roughness that fails under new light: albedo and relighting suffer most, novel view synthesis least.
$\mathcal{L}_{\text{sil}}$ mainly helps novel view synthesis, via boundary coverage.
$\mathcal{L}_{\text{cons}}$ affects image metrics least but keeps the surfels in a single thin layer (\cref{fig:ddist_ablation}), as oriented-point reconstruction expects.
A sphere initialization lowers every metric, most of all novel view synthesis and normals: surfels must reach the surface by gradient descent alone and can settle in local minima, notably on thin Ficus branches and leaves (\cref{sec:limitations}).

\begin{figure}[t]
  \centering
  \begin{subfigure}[t]{0.325\linewidth}
    \includegraphics[width=\linewidth, trim=112 25 0 0, clip]{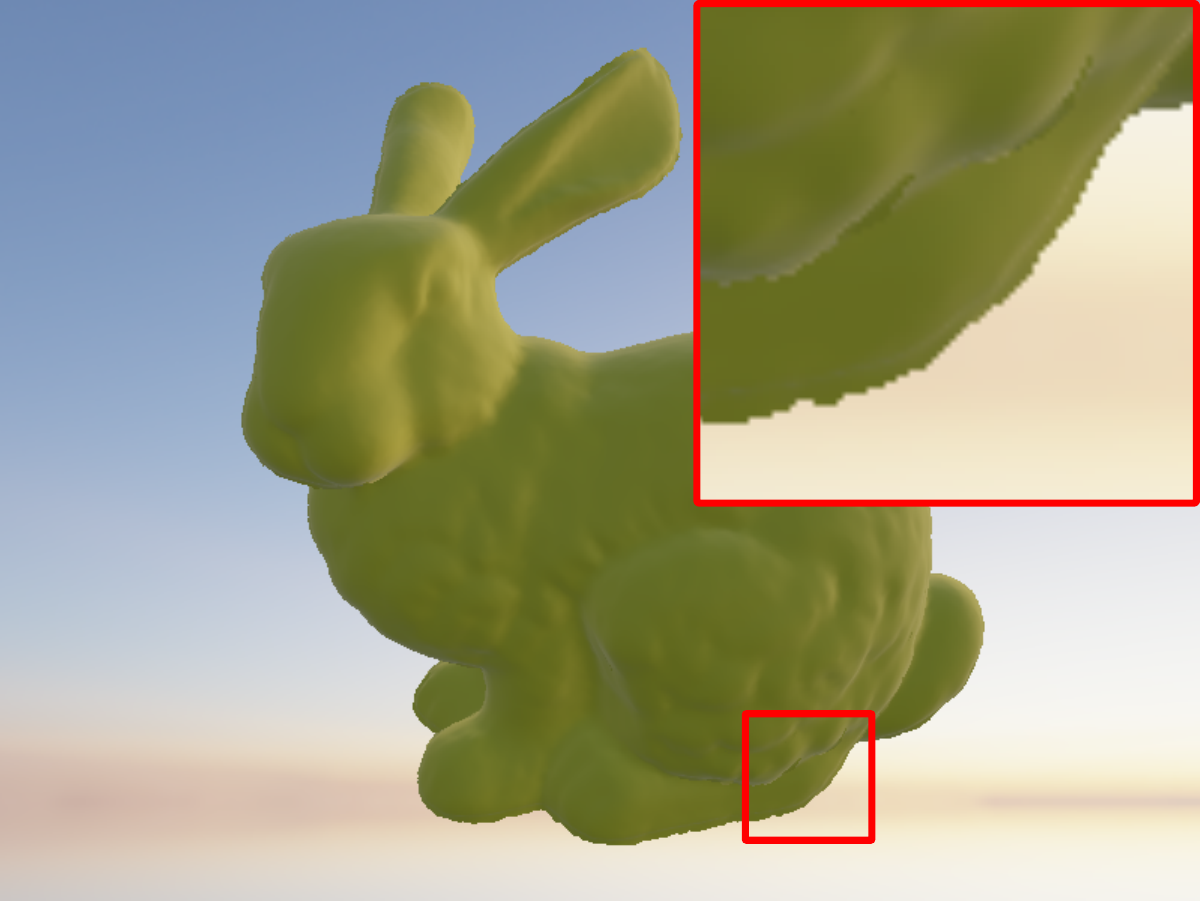}
    \caption{Ternary}
    \label{fig:grouping_ternary_noalpha}
  \end{subfigure}\hfill
  \begin{subfigure}[t]{0.325\linewidth}
    \includegraphics[width=\linewidth, trim=112 25 0 0, clip]{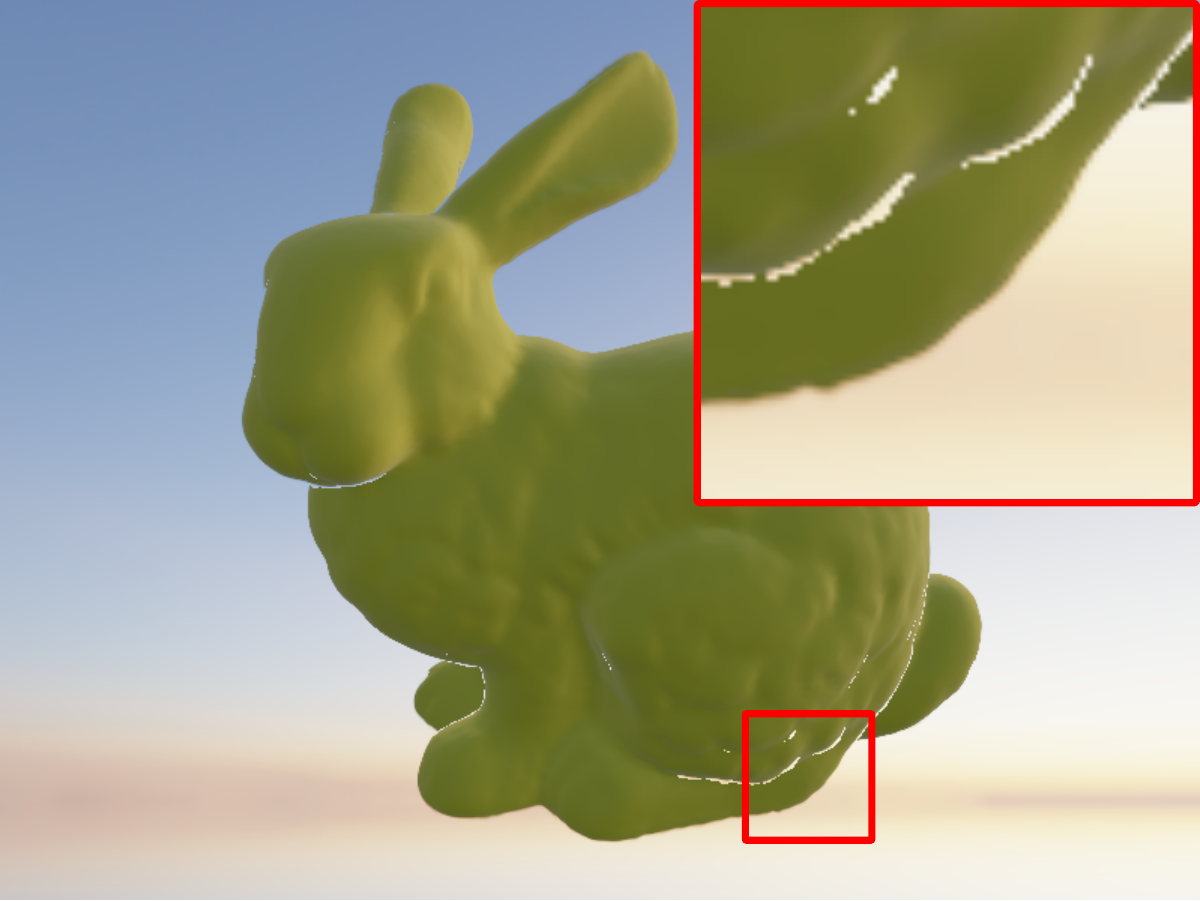}
    \caption{Ternary + Coverage}
    \label{fig:grouping_ternary_alpha}
  \end{subfigure}\hfill
  \begin{subfigure}[t]{0.325\linewidth}
    \includegraphics[width=\linewidth, trim=112 25 0 0, clip]{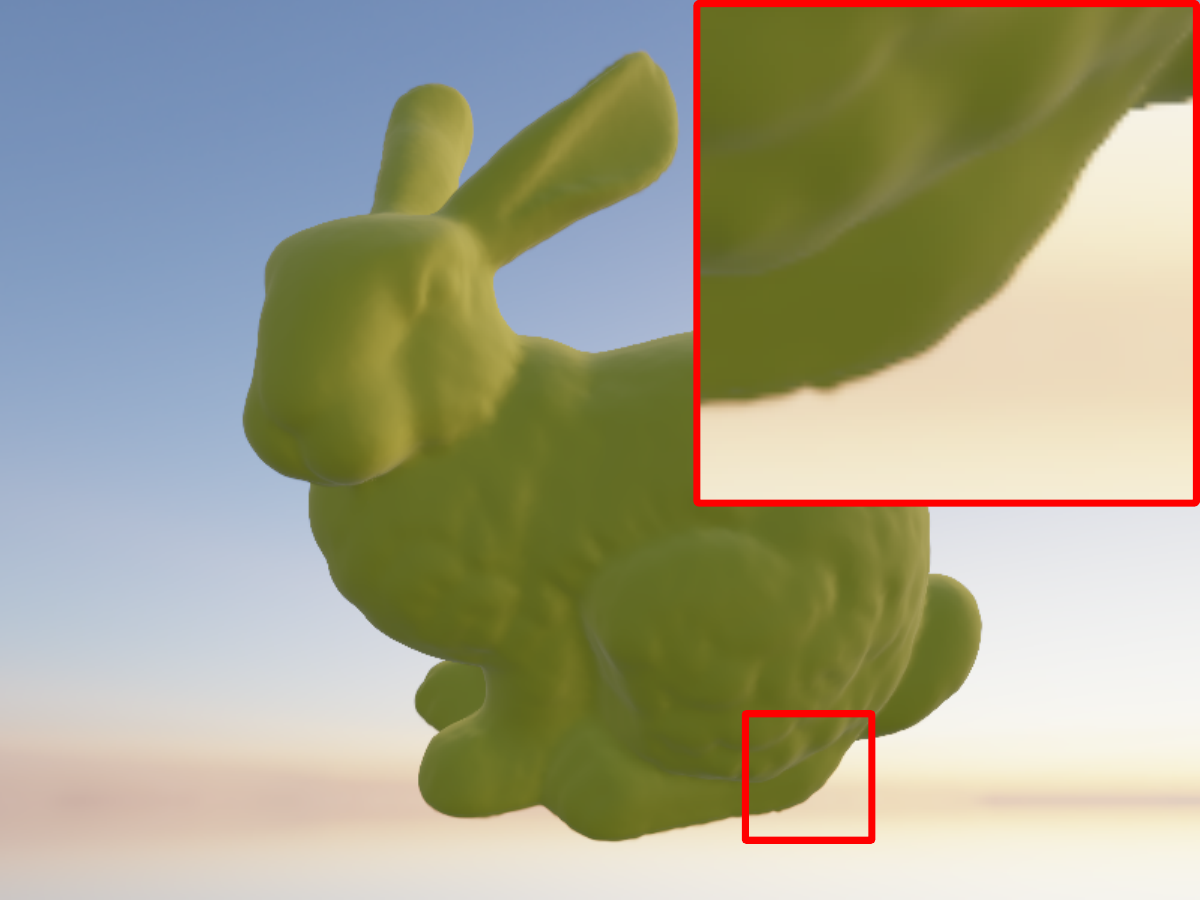}
    \caption{Ours}
    \label{fig:grouping_ours}
  \end{subfigure}\\[2pt]
  \begin{subfigure}[t]{0.37\linewidth}
    \includegraphics[width=\linewidth, trim=40 41 0 0, clip]{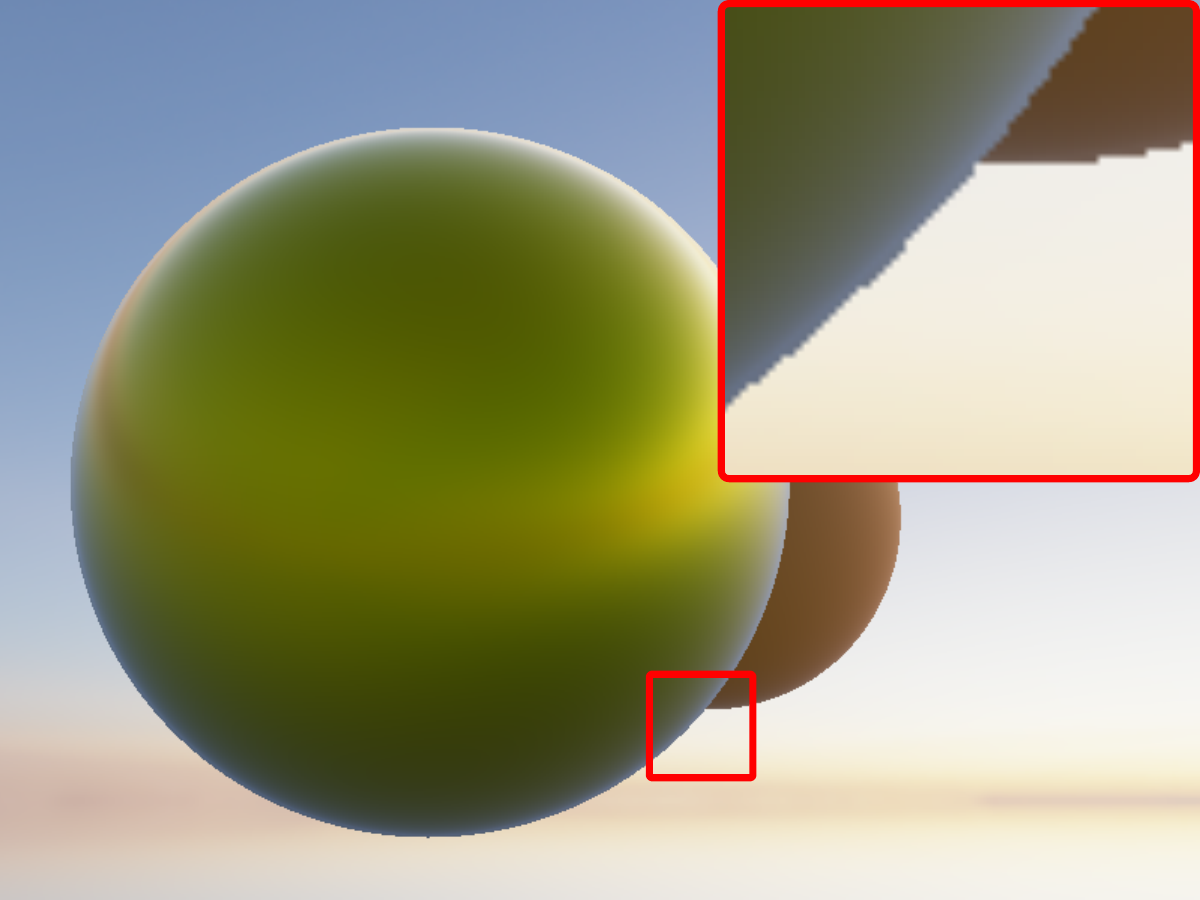}
    \caption{Ternary}
    \label{fig:compositing_ternary}
  \end{subfigure}\hspace{0.02\linewidth}
  \begin{subfigure}[t]{0.37\linewidth}
    \includegraphics[width=\linewidth, trim=40 41 0 0, clip]{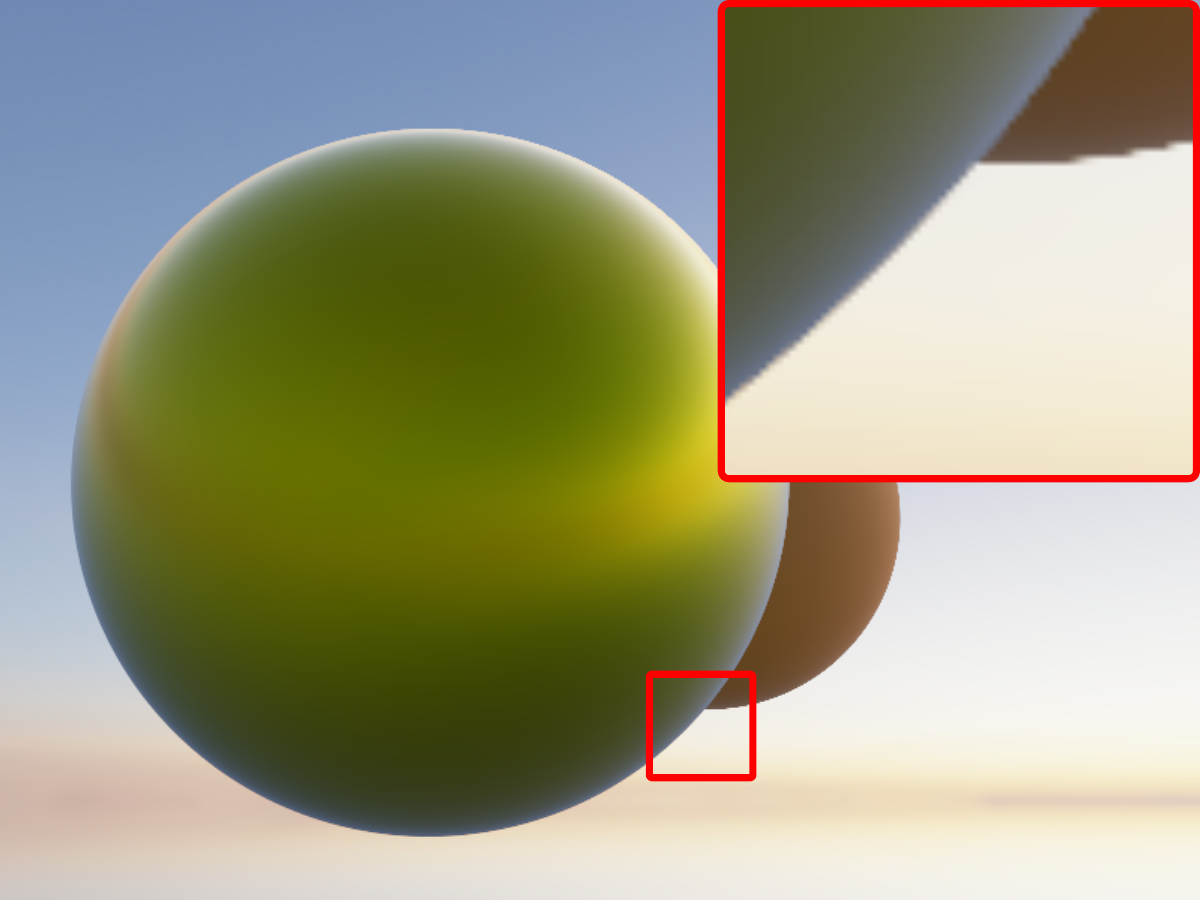}
    \caption{Ours}
    \label{fig:compositing_ours}
  \end{subfigure}
  \caption{\textbf{Grouping and coverage}: ternary depth test without (a, d) and with (b) coverage opacity, and ours (c, e).}
  \label{fig:ablation_grouping}
  \label{fig:compositing_ablation}
\end{figure}

\paragraph{Depth-interval grouping and coverage.}
\label{sec:ablation_grouping}
\label{sec:ablation_compositing}
On a point-sampled Stanford Bunny (\cref{fig:ablation_grouping}), the single-layer ternary depth test~\cite{weyrich2007hardware} blends all fragments within an $\varepsilon$-band of the stored depth into a fully opaque output, aliasing the silhouettes and causing self-occlusion errors where the haunch folds over the body (insets)~\subref{fig:grouping_ternary_noalpha}.
Adding coverage opacity smooths the silhouettes, but deeper surfels wrongly rejected by the test leave holes~\subref{fig:grouping_ternary_alpha}.
On two overlapping spheres, the ternary test is jagged at both silhouettes and where they overlap~\subref{fig:compositing_ternary}.
Our renderer avoids all these artifacts~\subref{fig:grouping_ours} and~\subref{fig:compositing_ours}, blending the overlap through the residual transmittance.

\begin{figure}[t]
  \centering
  \newcommand{\MIPimg}[1]{\figimg[width=\linewidth, trim=0 291 122 0, clip]{Figures/Ablations/mip_filtering/zoom3/#1.pdf}}
  \begin{subfigure}[t]{0.49\linewidth}
    \MIPimg{mip}
    \caption{Center-Precomputed (50\,ms)}
    \label{fig:mip_ours}
  \end{subfigure}\hfill
  \begin{subfigure}[t]{0.49\linewidth}
    \MIPimg{mipperpix}
    \caption{Per-Pixel Jacobian (65\,ms)}
    \label{fig:mip_perpixel}
  \end{subfigure}\\[2pt]
  \begin{subfigure}[t]{0.49\linewidth}
    \MIPimg{womip}
    \caption{No MIP Filter (42\,ms)}
    \label{fig:mip_none}
  \end{subfigure}\hfill
  \begin{subfigure}[t]{0.49\linewidth}
    \MIPimg{womip4x}
    \caption{$4{\times}$ Supersampling (87\,ms)}
    \label{fig:mip_4xss}
  \end{subfigure}
  \caption{\textbf{Object-space anti-aliasing}: horizon of a checkerboard plane of 4M surfels at $800{\times}600$, with frame times.}
  \label{fig:mip_ablation}
\end{figure}

\paragraph{Object-space anti-aliasing.}
\label{sec:ablation_mip}
Without filtering, a checkerboard plane shows moir\'{e} patterns toward the horizon (\cref{fig:mip_ablation}).
Our center-precomputed MIP filter removes them as cleanly as $4{\times}$ supersampling and the per-pixel Jacobian evaluation of AA-2DGS~\cite{younes2025anti}, at under 20\% overhead against about 55\% for the latter.

\section{Limitations and Conclusion}
\label{sec:limitations}

Our depth-interval grouping can merge distinct surfaces that lie closer together than the kernel extent, such as thin shells or near-parallel layers.
Moreover, the layer assignment depends on the surfel parameters, but the backward pass holds it fixed: the gradient never proposes moving a surfel to another layer, merging two layers, or splitting one.
The optimization can thus be trapped in local minima, which explains why 3DSS can struggle with very complex scenes and thin structures, and why it benefits from a good initialization (\cref{tab:tensoir_ablation}).
Finally, 3DSS assumes opaque surfaces; transparency would require a per-surfel attribute modulating the coverage opacity $\alpha_k$.

\paragraph{Conclusion.}
\label{sec:conc}
3DSS brings EWA surface splatting to physically-based inverse rendering: its surfels need no connectivity, like points, yet render as opaque, anti-aliased surfaces, like meshes, while remaining differentiable through visibility.
On Stanford-ORB and TensoIR Synthetic, it outperforms state-of-the-art mesh-based and Gaussian-based methods in novel view synthesis and relighting, with geometry on par with the strongest baselines, while recovering an explicit surface whose optimized surfels convert directly into meshes.
We hope that 3DSS renews interest in surface splatting as a rendering framework for inverse problems, complementing mesh-based and volumetric representations.

\phantomsection\pdfbookmark[0]{References}{references}
{
    \small
    \bibliographystyle{ieeenat_fullname}
    \bibliography{main}
}

\clearpage
\appendix
\maketitlesupplementary
\pdfbookmark[0]{Supplementary Material}{supplementary}

\crefalias{section}{appendix}
\crefalias{subsection}{subappendix}

These appendices provide the derivations and implementation details needed to reproduce 3D Surface Splatting (3DSS), together with additional analyses and results.
They follow the order of \cref{sec:method,sec:experiments}.
\Cref{sec:supp_rasterizer} details the surfel parameterization and the differentiable rasterizer, \cref{sec:supp_shading} the shading models of the two optimization stages, and \cref{sec:supp_tracing} the tracing of secondary rays through surfels.
\Cref{sec:supp_optimization} describes the optimization: initialization, adaptive density control, losses, and training schedule.
\Cref{sec:supp_setup} gives the experimental configuration, and \cref{sec:performance} analyzes the rendering performance.
\Cref{sec:supp_meshes} describes the conversions between surfels and triangle meshes shown in \cref{fig:teaser}, including the baking of the optimized materials into textured meshes.
Finally, \cref{sec:supp_ablation} extends the discussion of the ablation study, and \cref{sec:supp_qualitative} provides additional qualitative comparisons.

\section{Surfels and Differentiable Rasterization}
\label{sec:supp_rasterizer}

\begin{figure*}[t]
  \centering
  \includegraphics[width=0.99\textwidth]{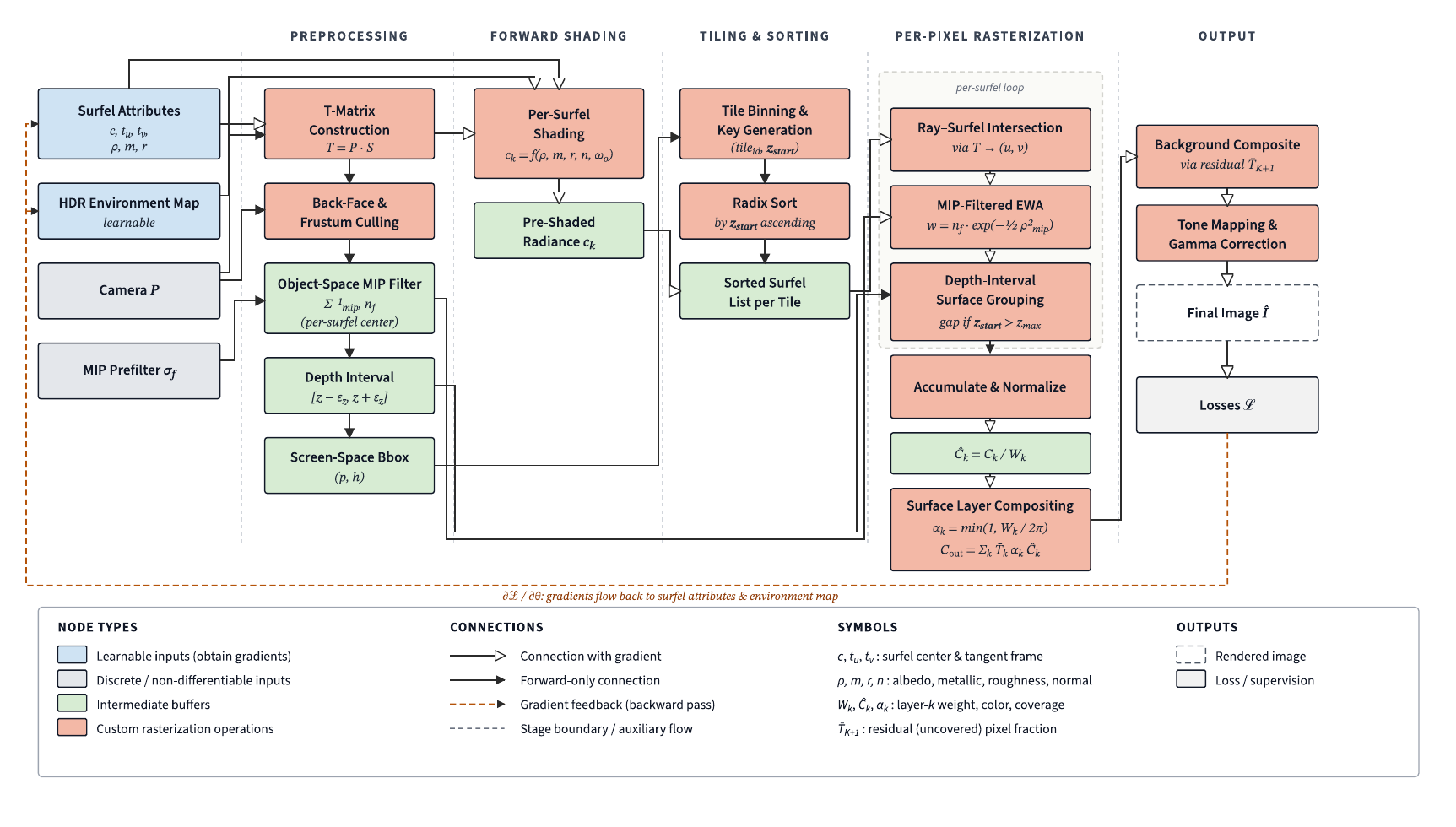}
  \caption{\textbf{Rendering pipeline of 3DSS.} Custom operations are shown in red, intermediate buffers in green, and learnable or discrete inputs on the left.
  \textbf{(1)~Preprocessing.} Each surfel is transformed to camera space, and its T-matrix is constructed for the analytic ray--surfel intersection and the screen-space bounding box (\cref{sec:supp_tmatrix}).
  \textbf{(2)~Forward shading.} Every surfel is shaded \emph{before} rasterization by a shading function of its attributes, the view direction, and the learnable environment map, which yields a pre-shaded linear radiance $\mathbf{c}_i$; the rasterizer is agnostic to the choice of this function (\cref{sec:supp_shading}).
  \textbf{(3)~Tiling and sorting.} Surfels are binned into screen-space tiles and radix-sorted by the start $z_{\text{start}}$ of their depth interval.
  \textbf{(4)~Per-pixel rasterization.} For each pixel, the viewing ray is intersected with each surfel, a MIP-filtered EWA kernel weight $w$ is evaluated at the hit, overlapping depth intervals are merged into surface layers, and the weighted contributions of each layer are normalized (\cref{alg:interval_grouping}). The accumulated weight $W_k$ of each layer gives its coverage opacity $\alpha_k = \min(1, W_k/2\pi)$, which drives the front-to-back compositing of all layers.
  \textbf{(5)~Output.} The composited HDR image is blended with the background, then tone-mapped and gamma-corrected. Gradients flow back through all operations to the surfel parameters and the environment map (\cref{sec:backward}).}
  \label{fig:pipeline}
\end{figure*}

\Cref{fig:pipeline} summarizes the rendering pipeline of 3DSS.
This appendix details the parameterization of a surfel (\cref{sec:parameterization}), the ray--surfel intersection and bounding box computed from its T-matrix (\cref{sec:supp_tmatrix}), the per-pixel grouping and compositing of surface layers (\cref{sec:supp_grouping}), and the backward pass through them (\cref{sec:backward}).

\subsection{Surfel Parameterization}
\label{sec:parameterization}

A surfel stores a center $\mathbf{c}$, a unit quaternion $\mathbf{q}$, two in-plane scales $(s_1, s_2)$, an albedo $\boldsymbol{\rho}$, a roughness $r$, a metallic $m$, and the coefficients of a degree-3 spherical-harmonics (SH) radiance function.
Its tangent vectors are $\mathbf{t}_u = s_1\,\mathbf{R}(\mathbf{q})\,\mathbf{e}_x$ and $\mathbf{t}_v = s_2\,\mathbf{R}(\mathbf{q})\,\mathbf{e}_y$, where $\mathbf{R}(\mathbf{q})$ is the rotation of $\mathbf{q}$, and its normal is $\mathbf{n} = \mathbf{R}(\mathbf{q})\,\mathbf{e}_z$.
The object space of the surfel maps the coordinates $(u, v)$ to $\mathbf{c} + u\,\mathbf{t}_u + v\,\mathbf{t}_v$, and its reconstruction kernel is the unit-variance isotropic Gaussian $\exp(-\frac12(u^2+v^2))$ of this space, cut off at $u^2 + v^2 < r_{\text{cut}}^2$ (\cref{eq:kernel}).
The support of a surfel is therefore the elliptical disk
\begin{equation}
  \mathcal{D} = \{\mathbf{c} + u\,\mathbf{t}_u + v\,\mathbf{t}_v \;:\; u^2 + v^2 \le r_{\text{cut}}^2\}\,,
  \label{eq:disk}
\end{equation}
and all of its anisotropy is carried by the tangent vectors.
We use $r_{\text{cut}} = 3$ throughout.

\paragraph{Material activations.}
Materials are stored as unconstrained logits and mapped to their physical ranges,
\begin{equation}
  \begin{gathered}
  \boldsymbol{\rho} = \rho_{\min} + (\rho_{\max} - \rho_{\min})\,\sigma(\tilde{\boldsymbol{\rho}}),\\
  r = r_{\max}\,\sigma(\tilde{r}), \qquad m = \sigma(\tilde{m}),
  \end{gathered}
  \label{eq:material_act}
\end{equation}
with $\sigma$ the logistic function.

\paragraph{Bounded-anisotropy scales.}
\label{sec:scale_param}
A surfel is a local sample of the surface, and its reconstruction kernel is expected to have a compact, nearly isotropic support.
If the two scales are optimized as independent log-scale variables, nothing prevents the optimizer from driving a surfel toward extreme anisotropy, $s_1 \gg s_2$, in which a single primitive covers an elongated region as a thin needle.
Such a kernel violates the local support assumed by the Shepard normalization (\cref{eq:shepard}) and by the coverage model (\cref{sec:compositing}), and destabilizes the optimization.
We therefore bound the anisotropy ratio $s_1/s_2$ by construction, through the parameterization, rather than with a penalty term or a clamp.

Since the bound applies to a ratio, we work in log scale and factor the scales into a size and a shape,
\begin{equation}
  s_1 = \mu\,e^{+d}, \qquad s_2 = \mu\,e^{-d},
  \label{eq:scale_split}
\end{equation}
where $\mu = \sqrt{s_1 s_2}$ is the geometric-mean radius and $s_1/s_2 = e^{2d}$ is the anisotropy ratio.
The constraint $s_1/s_2 \in [1/\kappa,\, \kappa]$ is then equivalent to $|d| \le \tfrac12\log\kappa$, which we satisfy by mapping an unconstrained shape variable $\delta \in \mathbb{R}$ through a hyperbolic tangent,
\begin{equation}
  d = \tfrac12\log\kappa \cdot \tanh\delta
  \quad\Longrightarrow\quad
  \frac{s_1}{s_2} = \kappa^{\tanh\delta} \in \left(\tfrac1\kappa,\ \kappa\right).
  \label{eq:scale_shape}
\end{equation}
With the size parameterized as $\mu = e^{\xi}$, the optimized variables are the unconstrained pair $(\xi, \delta)$, and
\begin{equation}
  s_1 = e^{\xi}\,\kappa^{+\frac12\tanh\delta}, \qquad
  s_2 = e^{\xi}\,\kappa^{-\frac12\tanh\delta}.
  \label{eq:scale_final}
\end{equation}
An isotropic kernel corresponds to $\delta = 0$, and the anisotropy ratio approaches the bound $\kappa$ only asymptotically, as $|\delta| \to \infty$.
Since size and shape are controlled by separate variables, the size of a surfel can change without interacting with the constraint on its shape.
In contrast to a penalty or a log-barrier on the ratio, the mapping introduces no additional loss weight; in contrast to clamping, it is smooth everywhere; and the gradient with respect to $\delta$ is not attenuated at the isotropic initialization, since $\tanh'(0) = 1$.

Surfels are initialized isotropically, with $\delta = 0$ (\cref{sec:supp_init}).
A given scale pair, such as one obtained by sampling a mesh (\cref{sec:mesh_sampling}), is converted by inverting \cref{eq:scale_final}: $\xi = \tfrac12\log(s_1 s_2)$ and $\tanh\delta = \log(s_1/s_2)/\log\kappa$, clamped to $(-1, 1)$ so that the initial value satisfies the bound strictly.
The splitting of adaptive density control (\cref{sec:density_control}) shrinks only the larger of the two scales and therefore preserves the bound.
We set $\kappa = 5$ in all experiments.

\subsection{Ray--Surfel Intersection and Bounding Box}
\label{sec:supp_tmatrix}

Following Weyrich et al.~\cite{weyrich2007hardware} and 2DGS~\cite{huang20242d}, the rasterizer derives both the ray--surfel intersection and the screen-space bounding box from the T-matrix of the surfel.
Let $\mathbf{t}'_u$, $\mathbf{t}'_v$, and $\mathbf{c}'$ be the tangent vectors and the center in camera space.
The homogeneous matrix $\mathbf{S} = [\,\mathbf{t}'_u\;\mathbf{t}'_v\;\mathbf{0}\;\mathbf{c}'\,]$, whose last row is $(0, 0, 0, 1)$, maps object-space points $(u, v, 0, 1)^\top$ to camera space, and the T-matrix $\mathbf{T} = \mathbf{P}\mathbf{S}$ maps them to clip space, where $\mathbf{P}$ is the projection matrix with the viewport transform folded in, so that $(x, y)$ below are pixel coordinates.

\paragraph{Intersection.}
The viewing ray of pixel $(x, y)$ is the intersection of the two clip-space planes $\mathbf{h}_x = (-1, 0, 0, x)^\top$ and $\mathbf{h}_y = (0, -1, 0, y)^\top$.
In object space, these planes become $\mathbf{h}_u = \mathbf{T}^\top\mathbf{h}_x$ and $\mathbf{h}_v = \mathbf{T}^\top\mathbf{h}_y$, and the hit point $(u, v)$ satisfies the two linear equations $h^1 u + h^2 v + h^4 = 0$ for $\mathbf{h} \in \{\mathbf{h}_u, \mathbf{h}_v\}$, where superscripts index vector components.
Their solution is
\begin{equation}
  u = \frac{h_u^2 h_v^4 - h_u^4 h_v^2}{h_u^1 h_v^2 - h_u^2 h_v^1}\,,\qquad
  v = \frac{h_u^4 h_v^1 - h_u^1 h_v^4}{h_u^1 h_v^2 - h_u^2 h_v^1}\,,
  \label{eq:uv}
\end{equation}
which is differentiable in the entries of $\mathbf{T}$ and hence in the surfel parameters.
The kernel weight of \cref{eq:mip_weight} is evaluated at this point.

\paragraph{Bounding box.}
In homogeneous object coordinates $(u, v, 1)$, the unit circle is the conic $\mathbf{M} = \operatorname{diag}(1, 1, -1)$.
Let $\mathbf{T}_1, \mathbf{T}_2, \mathbf{T}_4 \in \mathbb{R}^3$ be the rows of $\mathbf{T}$ that produce the clip coordinates $x$, $y$, and $w$, restricted to the columns of $u$, $v$, and the center.
The screen-space bounding rectangle of the projected circle has center $(p_1, p_2)$ and half-extents $(h_1, h_2)$, with~\cite{sigg2006gpu,huang20242d}
\begin{equation}
  p_i = \frac{\langle \mathbf{T}_i, \mathbf{M}\mathbf{T}_4\rangle}{\langle \mathbf{T}_4, \mathbf{M}\mathbf{T}_4\rangle}\,,\qquad
  h_i = \sqrt{p_i^2 - \frac{\langle \mathbf{T}_i, \mathbf{M}\mathbf{T}_i\rangle}{\langle \mathbf{T}_4, \mathbf{M}\mathbf{T}_4\rangle}}\,,
  \label{eq:bbox}
\end{equation}
for $i \in \{1, 2\}$; the half-extents are then scaled by $r_{\text{cut}}$ for tile binning.
Since $\mathbf{M}$ is its own inverse, these expressions only involve the projected dual conic $\mathbf{T}\mathbf{M}\mathbf{T}^\top$, and $\mathbf{T}$ itself is never inverted.
This matters in our setting: the restricted T-matrix becomes singular when a surfel is seen edge-on or when one of its scales becomes very small, which happens routinely during optimization.

\subsection{Depth-Interval Grouping and Compositing}
\label{sec:supp_grouping}

\begin{algorithm}[t]
\caption{Per-pixel rasterization with depth-interval grouping and coverage-based compositing.}
\label{alg:interval_grouping}
\begin{algorithmic}[1]
\Require Surfels $\mathcal{S}$ of the tile, sorted by $z_{\text{start}}$ in ascending order; pixel $(x, y)$
\Ensure Composited color $\mathbf{C}_{\text{out}}$ and residual transmittance $\bar{T}$
\State $\mathbf{C}_{\text{out}} \gets \mathbf{0}$, $\bar{T} \gets 1$, $K \gets 0$ \Comment{Output}
\State $\mathbf{C} \gets \mathbf{0}$, $W \gets 0$, $z^{\max}_{\text{end}} \gets -\infty$ \Comment{Current layer}
\For{each surfel $i \in \mathcal{S}$}
    \State $(u, v) \gets$ ray--surfel intersection (\cref{eq:uv})
    \State $\rho^2_{\text{mip}} \gets \mathbf{u}^\top \boldsymbol{\Sigma}^{-1}_{\text{mip},i}\,\mathbf{u}$, \quad $\mathbf{u} = (u, v)^\top$
    \If{$\rho^2_{\text{mip}} \geq r_{\text{cut}}^2$}
        \State \textbf{continue} \Comment{Surfel does not cover the pixel}
    \EndIf
    \State $w \gets n_{f,i} \exp(-\tfrac{1}{2}\,\rho^2_{\text{mip}})$
    \If{$W > 0$ \textbf{and} $z_{\text{start},i} > z^{\max}_{\text{end}}$} \Comment{Gap}
        \State \Call{FlushLayer}{}
        \If{$K = K_{\max}$} \textbf{break} \EndIf
        \State $\mathbf{C} \gets \mathbf{0}$, $W \gets 0$, $z^{\max}_{\text{end}} \gets -\infty$
    \EndIf
    \State $z^{\max}_{\text{end}} \gets \max(z^{\max}_{\text{end}},\; z_{\text{end},i})$
    \State $\mathbf{C} \gets \mathbf{C} + w\,\mathbf{c}_i$, \quad $W \gets W + w$
\EndFor
\If{$W > 0$ \textbf{and} $K < K_{\max}$} \Call{FlushLayer}{} \EndIf
\Statex
\Procedure{FlushLayer}{}
    \State $\alpha \gets \min(1,\, W/2\pi)$ \Comment{Coverage opacity}
    \State $\mathbf{C}_{\text{out}} \gets \mathbf{C}_{\text{out}} + \bar{T}\,\alpha\,\mathbf{C}/W$ \Comment{Shepard-normalized color}
    \State $\bar{T} \gets \bar{T}\,(1 - \alpha)$, \quad $K \gets K + 1$
\EndProcedure
\end{algorithmic}
\end{algorithm}

\Cref{alg:interval_grouping} gives the per-pixel loop of the rasterizer, which implements the depth-interval grouping of \cref{sec:depth_grouping}, the coverage opacity $\alpha_k = \min(1, W_k/2\pi)$ of \cref{eq:alpha}, and the front-to-back compositing of \cref{eq:compositing}.
The depth interval $[z_{\text{start}}, z_{\text{end}}] = [z - \varepsilon_z,\, z + \varepsilon_z]$ of each surfel is computed once per frame, in preprocessing, from the camera-space depth $z$ of its center and the depth extent $\varepsilon_z = r_{\text{cut}}\,\|(t'_{u,z},\, t'_{v,z})\|$ of its support; it is the support function of the disk of \cref{eq:disk} along the optical axis (\cref{eq:support}).
The MIP filter of \cref{sec:mip_filtering} is precomputed at the center of each surfel, which stores the three unique entries of the symmetric matrix $\boldsymbol{\Sigma}^{-1}_{\text{mip}}$ and the amplitude $n_f$ alongside the rows of $\mathbf{T}$.
The cutoff radius $r_{\text{cut}}$ enters the pipeline in three places, which must agree: the bounding box used for tile binning (\cref{eq:bbox}), the depth extent $\varepsilon_z$ used for grouping, and the per-pixel acceptance test $\rho^2_{\text{mip}} < r_{\text{cut}}^2$.

Since the backward pass needs the statistics of every layer of a pixel (\cref{sec:backward}), the number of composited layers is capped at $K_{\max} = 16$: once the $K_{\max}$-th layer is flushed, the remaining surfels are discarded.
For the object-level scenes of our experiments, most pixels contain one to three layers, and the cap is reached only at complex depth discontinuities.

\subsection{Backward Pass}
\label{sec:backward}

For a pixel with layers $k = 1, \ldots, K$ and background color $\mathbf{C}_{\text{bg}}$, the output is $\mathbf{C}_{\text{out}} = \sum_k \bar{T}_k\,\alpha_k\,\hat{\mathbf{C}}_k + \bar{T}_{K+1}\,\mathbf{C}_{\text{bg}}$.
Let $\mathbf{B}_k$ be the color composited behind layer $k$, defined recursively from back to front by $\mathbf{B}_K = \mathbf{C}_{\text{bg}}$ and $\mathbf{B}_{k} = \alpha_{k+1}\hat{\mathbf{C}}_{k+1} + (1 - \alpha_{k+1})\,\mathbf{B}_{k+1}$.
Since $\mathbf{C}_{\text{out}} = \sum_{j<k} \bar{T}_j\,\alpha_j\,\hat{\mathbf{C}}_j + \bar{T}_k\,(\alpha_k\hat{\mathbf{C}}_k + (1 - \alpha_k)\,\mathbf{B}_k)$, where neither the sum nor $\bar{T}_k$ and $\mathbf{B}_k$ depend on layer $k$,
\begin{equation}
  \frac{\partial \mathbf{C}_{\text{out}}}{\partial \hat{\mathbf{C}}_k} = \bar{T}_k\,\alpha_k\,,\qquad
  \frac{\partial \mathbf{C}_{\text{out}}}{\partial \alpha_k} = \bar{T}_k\,\bigl(\hat{\mathbf{C}}_k - \mathbf{B}_k\bigr)\,,
  \label{eq:grad_layer}
\end{equation}
a form that involves no division by $1 - \alpha_k$ and therefore remains valid for fully covered layers, $\alpha_k = 1$, which the clamped coverage of \cref{eq:alpha} produces routinely.
Through the Shepard normalization and the clamped coverage, a surfel $i$ of layer $k$ with weight $w_i$ and color $\mathbf{c}_i$ receives
\begin{equation}
  \frac{\partial \hat{\mathbf{C}}_k}{\partial \mathbf{c}_i} = \frac{w_i}{W_k}\,,\quad
  \frac{\partial \hat{\mathbf{C}}_k}{\partial w_i} = \frac{\mathbf{c}_i - \hat{\mathbf{C}}_k}{W_k}\,,\quad
  \frac{\partial \alpha_k}{\partial w_i} = \frac{\mathbb{1}[W_k < 2\pi]}{2\pi}\,.
  \label{eq:grad_surfel}
\end{equation}
The two terms of the weight gradient have distinct roles.
The first changes the share of the surfel in the color of its layer, and vanishes when all surfels of the layer agree.
The second is the visibility gradient: it compares the layer with what lies behind it, and is nonzero only at partially covered pixels, along silhouettes and occlusion boundaries, where the clamp of \cref{eq:alpha} is inactive.
The weight gradient then propagates to the surfel parameters through the kernel of \cref{eq:mip_weight}, the intersection of \cref{eq:uv}, and the T-matrix.
The layer assignment itself is piecewise constant in the surfel parameters and is held fixed in the backward pass, as discussed in \cref{sec:limitations}.

\paragraph{Implementation.}
The forward and backward passes are custom CUDA kernels that follow the tile-based paradigm of 3DGS~\cite{kerbl2023gaussian}.
Tiles of $16 \times 16$ pixels are processed by individual thread blocks, which load the sorted surfels of their tile into shared memory in batches.
In the forward pass, the intersection, the MIP-filtered kernel evaluation, the grouping, the Shepard normalization, and the compositing are fused into a single kernel invocation per tile.
The backward kernel makes two passes over the sorted surfels of the tile: the first replays the gap detection of the forward pass to recover the per-layer statistics $(\hat{\mathbf{C}}_k, W_k)$, and the second assigns each surfel to its layer and distributes the gradients of \cref{eq:grad_layer,eq:grad_surfel} to the surfels, with atomic operations to resolve concurrent writes from different pixels.

\section{Shading Models}
\label{sec:supp_shading}

The rasterizer composites a pre-shaded radiance $\mathbf{c}_i$ per surfel and is agnostic to how it is computed (\cref{sec:forward_shading}).
The two optimization stages use different shading functions, because they solve different problems: the geometry stage needs a model that explains appearance well enough to recover the shape, and the inverse-rendering stage needs a physically-based model that separates material from illumination.

\subsection{Geometry Stage: Split-Sum Image-Based Lighting}
\label{sec:forward_shading_practice}

In the geometry stage, each training view is rendered twice: once with the SH radiance of each surfel evaluated along the view direction, whose degree is annealed upward to 3 during training, and once with the physically-based model below.
The outgoing radiance of the physically-based image at a surfel with albedo $\boldsymbol{\rho}$, roughness~$r$, metallic~$m$, unit normal~$\mathbf{n}$, and view direction $\boldsymbol{\omega}_o$ is
\begin{equation}
  \mathbf{c}
  = \underbrace{
      (1 - m)\,\boldsymbol{\rho}\,
      L_d(\mathbf{n})
    }_{\text{diffuse}}
  \;+\;
  \underbrace{
      L_s(\mathbf{r}, r)\,
      \bigl(\mathbf{F}_0\,\beta_1 + \beta_2\bigr)
    }_{\text{specular}}\,,
  \label{eq:ibl}
\end{equation}
where $L_d(\mathbf{n})$ is the irradiance queried from a pre-convolved diffuse environment map along the normal, $L_s(\mathbf{r}, r)$ is the GGX-prefiltered specular radiance queried along the reflection direction $\mathbf{r} = 2(\boldsymbol{\omega}_o \cdot \mathbf{n})\mathbf{n} - \boldsymbol{\omega}_o$ at a MIP level determined by the roughness, and $(\beta_1, \beta_2)$ are the scale and bias read from a precomputed BRDF integration lookup table indexed by $(\boldsymbol{\omega}_o \cdot \mathbf{n},\, r)$~\cite{karis2013real}.
The base reflectivity $\mathbf{F}_0 = 0.04\,(1-m) + m\,\boldsymbol{\rho}$ follows the metallic workflow of NVDiffRec~\cite{Munkberg_2022_CVPR}: it reduces to an achromatic dielectric reflectance for $m = 0$, and takes the color of the albedo for $m = 1$, where the diffuse lobe vanishes.
The environment is a learnable cubemap whose radiance is a softplus of the stored parameter, which keeps it positive; since the map is optimized, its prefiltered MIP pyramid is rebuilt at every iteration, as in NVDiffRec~\cite{Munkberg_2022_CVPR}.
This image constrains the geometry to also explain a microfacet BRDF, but its materials and cubemap serve only as a proxy: both are re-initialized at the start of the inverse-rendering stage.

\subsection{Inverse-Rendering Stage: Monte Carlo Shading}
\label{sec:supp_mc}

In the inverse-rendering stage, the radiance of each surfel is the Monte Carlo estimate of \cref{eq:mc_shading}, with the incident radiance $L = V L_{\text{env}} + L_{\text{ind}}$ of \cref{eq:incident}, whose visibility and indirect terms are obtained by tracing secondary rays (\cref{sec:supp_tracing}).

\paragraph{BRDF.}
The specular lobe is a GGX microfacet BRDF~\cite{walter2007microfacet} with the Smith--Schlick geometry term and the spherical-Gaussian Fresnel approximation of Karis~\cite{karis2013real},
\begin{equation}
  f_s = \frac{F(\boldsymbol{\omega}_o\!\cdot\!\mathbf{h})\;\alpha^2}
             {4\pi\,D^2\,G_o\,G_s}\,,\qquad
  \alpha = r^2,\quad
  k = \tfrac{1}{8}\,(r + 1)^2,
  \label{eq:ggx}
\end{equation}
where $\mathbf{h}$ is the half vector, $D = (\mathbf{n}\!\cdot\!\mathbf{h})^2(\alpha^2 - 1) + 1$ is the Trowbridge--Reitz denominator, and $G_o, G_s$ are the Smith terms $(\mathbf{n}\!\cdot\!\boldsymbol{\omega})(1-k) + k$ for the outgoing and incident directions.
\Cref{eq:ggx} is the product $D_{\text{GGX}}\,F\,G/(4\,(\mathbf{n}\cdot\boldsymbol{\omega}_o)(\mathbf{n}\cdot\boldsymbol{\omega}_s))$ in which the cosines of the Smith terms cancel those of the denominator.
The Fresnel term uses the same base reflectivity $\mathbf{F}_0$ as the geometry stage, and the diffuse lobe the same factor $(1-m)$.

\paragraph{Environment map.}
The environment $L_{\text{env}}$ is an equirectangular map stored as an unconstrained parameter under a positivity-ensuring activation: an exponential, or a softplus for the HDR captures of Stanford-ORB.
Its radiance is clamped from above, so that no single texel can grow without limit to explain a localized residual.

\paragraph{Sampling bias and the roughness ceiling.}
A single direction set serves both lobes of \cref{eq:mc_shading}, and this is the main limitation of our estimator.
A solid-angle-uniform set is well suited to the cosine-weighted diffuse integral but poorly matched to a narrow specular lobe: at a fixed budget, the specular term is estimated far less accurately than the diffuse one, and increasingly so as the roughness decreases.
The optimizer can reduce this estimation error simply by widening the lobe, and the roughness therefore acquires a systematic upward bias that no reweighting of the loss removes.
The natural remedy is to draw a second set of directions from the microfacet distribution and to combine the two sets with the balance heuristic~\cite{veach1995optimally}; the obstacle is that visibility and interreflection are cached for the traced direction set (\cref{sec:supp_caching}), so that a second set requires either a second trace or decoupling the shading directions from the cached ones.

\section{Secondary-Ray Tracing}
\label{sec:supp_tracing}

The visibility $V$ and the indirect radiance $L_{\text{ind}}$ of \cref{eq:incident} come from secondary rays, which we resolve with the same surface model as camera rays: a ray is intersected with the support disks of the surfels, the hits are grouped into surface layers by the overlap of their depth intervals, and the first layer beyond the originating surface determines $V$ and $L_{\text{ind}}$ through \cref{eq:trace}.
Two differences with the rasterizer require care.
A ray tracer does not receive the hits in depth order, so the grouping must be made independent of the order of the hits (\cref{sec:supp_merge}), and a secondary ray starts inside the layer of the shaded surfel, which must be excluded (\cref{sec:supp_exclusion}).

\subsection{Disk Supports, Intersection, and BVH}
\label{sec:supp_bvh}

\paragraph{One support formula.}
The tracer rests on a single identity.
For any vector $\mathbf{a}$, the support disk $\mathcal{D}$ of \cref{eq:disk} satisfies
\begin{equation}
  \max_{\mathbf{p} \in \mathcal{D}}\; (\mathbf{p} - \mathbf{c})\cdot\mathbf{a}
  = r_{\text{cut}}\,\sqrt{(\mathbf{t}_u\cdot\mathbf{a})^2 + (\mathbf{t}_v\cdot\mathbf{a})^2}\,,
  \label{eq:support}
\end{equation}
by the Cauchy--Schwarz inequality applied to $u\,(\mathbf{t}_u\cdot\mathbf{a}) + v\,(\mathbf{t}_v\cdot\mathbf{a})$ over $u^2 + v^2 \le r_{\text{cut}}^2$.
With $\mathbf{a}$ a world axis, \cref{eq:support} gives the tight axis-aligned bounding box of the disk; with $\mathbf{a}$ the optical axis of the camera, it gives the depth extent $\varepsilon_z$ of the rasterizer (\cref{sec:supp_grouping}); and with $\mathbf{a}$ the unit direction $\mathbf{d}$ of a ray with origin $\mathbf{o}$, it gives the half-width $\varepsilon$ of the interval $[t_{\text{start}}, t_{\text{end}}] = [t_c - \varepsilon,\, t_c + \varepsilon]$ that the disk occupies along the ray, where $t_c = (\mathbf{c} - \mathbf{o})\cdot\mathbf{d}$ is the projection of its center.
Bounding boxes, rasterizer depth extents, and ray depth extents are therefore one and the same formula.
Note that $t_c$ is not the hit distance unless the ray meets the disk face-on, but the hit distance always lies in the interval.

\paragraph{Intersection.}
The ray--disk test is the planar analogue of the ray--triangle test.
With the unnormalized normal $\mathbf{n} = \mathbf{t}_u \times \mathbf{t}_v$, the hit distance is $t = (\mathbf{c} - \mathbf{o})\cdot\mathbf{n} \,/\, (\mathbf{d}\cdot\mathbf{n})$; rays that meet the disk from behind ($\mathbf{d}\cdot\mathbf{n} > 0$) are rejected, matching the back-face culling of the rasterizer.
The local coordinates of the hit point $\mathbf{p} = \mathbf{o} + t\,\mathbf{d} - \mathbf{c}$ are $u = \mathbf{p}\cdot\mathbf{t}_u / \|\mathbf{t}_u\|^2$ and $v = \mathbf{p}\cdot\mathbf{t}_v / \|\mathbf{t}_v\|^2$, which is exact because the quaternion parameterization keeps $\mathbf{t}_u \perp \mathbf{t}_v$, and the weight is the kernel of \cref{eq:kernel}.
A hit counts only if it passes the cutoff $u^2 + v^2 < r_{\text{cut}}^2$, as in the rasterizer.
No MIP filter is applied: the prefilter of \cref{sec:mip_filtering} maps a pixel footprint into the tangent frame, and a secondary ray has no pixel footprint, so that $\boldsymbol{\Sigma}_{\text{mip}} = \mathbf{I}$ and $n_f = 1$.

Since the kernel is evaluated in the object space of each surfel, at the point where the ray crosses its plane, the accumulated weight $W$ of a layer is an intrinsic quantity of the surface: for coplanar surfels, all planes are crossed at the same point, and $W$ does not depend on the direction of the ray.
The coverage opacity $\alpha = \min(1, W/2\pi)$ of \cref{eq:alpha} therefore carries over to secondary rays of any direction; on curved surfaces, the crossing points of neighboring tangent planes drift apart toward grazing directions, where $W$ becomes noisier.

\paragraph{Bounding volume hierarchy.}
We build a linear BVH~\cite{karras2012maximizing} over the tight bounding boxes of the support disks given by \cref{eq:support}.
Since the geometry is frozen in the inverse-rendering stage, the BVH is built once; this takes about 20\,ms for 250K surfels.

\subsection{Order-Independent Interval Merging}
\label{sec:supp_merge}

The rasterizer forms surface layers by scanning the surfels in ascending order of their interval start, an order that its depth sort provides.
A BVH traversal visits the hits of a ray in an arbitrary order.
Sorting them per ray requires either a hit buffer as large as the deepest layer or several traversals that each collect the $k$ nearest remaining hits, as in ray tracers of Gaussian primitives~\cite{moenne2024gaussian,gu2024IRGS}; both are costly here.
We therefore characterize a layer by a condition that does not depend on the order in which its hits are visited, and that needs only constant state per ray.

Let $L$ be the start of a layer, \ie, the smallest interval start among the hits that do not belong to a previous layer, and define, for $E \ge L$,
\begin{equation}
  F(E) = \max\bigl(\{E\} \cup \{\, t_{\text{end},k} \;:\; L \le t_{\text{start},k} \le E \,\}\bigr)\,,
  \label{eq:closure}
\end{equation}
the farthest end among the hits that start between $L$ and $E$, or $E$ itself if no such hit extends beyond it; only hits that pass the back-face and cutoff tests enter the set.
$F$ is non-decreasing and satisfies $F(E) \ge E$.
The layer produced by the ascending scan is the set $\{k : L \le t_{\text{start},k} \le E^*\}$, where $E^*$ is the smallest fixed point of $F$ not below $L$.
Indeed, the scan closes the layer at the first start beyond its running maximum end $E_{\text{scan}}$, so that every member starts at or before $E_{\text{scan}}$, every other hit starts after it, and $F(E_{\text{scan}}) = E_{\text{scan}}$.
Conversely, the member intervals cover $[L, E_{\text{scan}}]$ without gaps, so that every $E \in [L, E_{\text{scan}})$ is covered by a member interval that starts at or before $E$ and ends after it; hence $F(E) > E$, and no smaller fixed point exists.
The iteration $E \leftarrow F(E)$, started at $E = L$, therefore increases strictly while $E < E^*$; it never overshoots, since $E \le E^*$ implies $F(E) \le F(E^*) = E^*$; and it terminates, since $F$ takes finitely many values.
The tracer thus obtains, without sorting, the layers that the ascending scan of the rasterizer would produce along the ray.

Each evaluation of $F$ is a maximum over the hits, which does not depend on their order and is computed in a single BVH traversal.
The same traversal also returns, at no extra cost, the smallest interval start greater than $E$, which is the start of the next layer once $E = E^*$.
Besides the usual ray--box test, a BVH node is skipped when the projection $[t_{\text{lo}}, t_{\text{hi}}]$ of its bounding box onto the ray satisfies $t_{\text{hi}} < L$, since every interval below the node then ends before the layer starts, or when $t_{\text{lo}}$ exceeds both $E$ and the smallest start beyond $E$ found so far, since no interval below the node can then contribute to $F(E)$ or lower that start.
Both tests are conservative, because the support disk of every surfel lies inside the bounding box of each of its ancestor nodes, so that $t_{\text{lo}} \le t_{\text{start}}$ and $t_{\text{end}} \le t_{\text{hi}}$ for every interval below the node.
In our experiments, each fixed-point iteration converges in at most 11 steps; we cap it at 32 steps, a bound that was never reached.

\subsection{Excluding the Originating Surface}
\label{sec:supp_exclusion}

A secondary ray starts at the center of the shaded surfel, and therefore inside the surface layer of that surfel, which must not occlude it.
We exclude this layer with a first fixed-point iteration, seeded with $L = -\infty$ and $E = \varepsilon_0$, the half-width of the interval of the shaded surfel along the ray (\cref{eq:support} with $\mathbf{a} = \mathbf{d}$).
Setting $L = -\infty$ ensures that this iteration also absorbs the surfels whose intervals start before the origin but extend beyond it, such as large neighboring disks that straddle the origin; surfels that lie entirely behind the origin end before $\varepsilon_0$ and do not affect it.
This first iteration accumulates nothing and returns the start of the next layer.
A second iteration, started there, accumulates the kernel weights and SH radiances of that layer, from which its coverage $\alpha$ and Shepard-normalized radiance $\hat{\mathbf{C}}$, and hence $V = 1 - \alpha$ and $L_{\text{ind}} = \alpha\,\hat{\mathbf{C}}$ (\cref{eq:trace}), follow.
A ray along which no further layer is found escapes to the environment, with $V = 1$ and $L_{\text{ind}} = 0$.
The SH radiance of each hit surfel is evaluated in the direction $-\mathbf{d}$, toward the shaded surfel.

\paragraph{Approximations.}
Three approximations follow from this design and are worth stating explicitly.
First, only the first layer is traced, and its residual $1 - \alpha$ is treated as unoccluded.
In the interior of a surface, the layer is fully covered ($\alpha = 1$) and the visibility is binary, but at the partially covered edges of an occluder, a fraction of the environment light reaches the shaded surfel even when another surface lies behind the edge.
Second, the exclusion absorbs every surface whose interval overlaps that of the originating layer, so that a surface closer to the origin than about one support diameter along the ray casts no shadow: self-shadowing at scales below the kernel support, such as contact shadows in tight creases, is not resolved.
Third, back faces are rejected as in the rasterizer, which is correct for closed, consistently oriented surfaces but lets light pass through open or one-sided geometry; this is one reason why thin sheets are represented by two oppositely oriented layers of surfels (\cref{sec:supp_setup}).

\subsection{Cost and Caching}
\label{sec:supp_caching}

Each GPU thread traces one ray, and the $S$ rays of a surfel are assigned to consecutive threads, so that the threads of a warp share the same origin and visit largely the same BVH nodes.
A ray requires about five BVH traversals on average over the two fixed-point iterations, and the trace costs about $0.16\,\mu$s per ray, \ie, 2.5\,s for 250K surfels at $S = 64$.
Since the geometry is frozen in the inverse-rendering stage, the geometric outcome of the trace does not change during that stage: we trace once, at the stage boundary, and cache the geometric result of every ray.
The $S$ directions of each surfel are therefore fixed for the whole stage: its Fibonacci lattice is randomly rotated once, when the rays are traced.
Only the radiance of the hit surfels is re-evaluated at each iteration, from their SH functions, which keep training (\cref{sec:optimization}).

\section{Optimization}
\label{sec:supp_optimization}

This appendix details the components of the optimization summarized in \cref{sec:optimization}: initialization (\cref{sec:supp_init}), adaptive density control (\cref{sec:density_control}), the losses (\cref{sec:supp_losses}), and the training schedule (\cref{sec:training_schedule}).

\subsection{Initialization}
\label{sec:supp_init}

We initialize the surfels from the dense depth maps that a monocular depth estimator, Depth Anything~3~\cite{lin2026depth}, predicts for the training views.
The depth maps are unprojected into 3D with the known camera intrinsics and poses, which yields a dense but noisy point cloud; on Stanford-ORB, this gives about 100K initial surfels per scene.
Normals are estimated from the spatial gradients of the depth maps, and the rotation of each surfel is the quaternion that aligns its local $z$-axis with the estimated normal.
The initial scales are isotropic ($\delta = 0$ in \cref{eq:scale_final}), set to the mean distance to the orientation-aware nearest neighbors multiplied by $\eta = 1.5$, so that the initial kernel supports overlap and approximate the partition of unity on which the coverage model relies (\cref{sec:compositing}).
The albedo is initialized from the point colors when available, the roughness to $0.1$, and the cubemap of the geometry stage to a uniform low-energy state.
Alternatively, 3DSS can be initialized without a depth prior, from surfels sampled uniformly on a sphere enclosing the object; this configuration is evaluated in the ablation study (\cref{sec:initialization_exp}).

\subsection{Adaptive Density Control}
\label{sec:density_control}

The coverage model of \cref{sec:compositing} assumes that the surfel kernels locally approximate a partition of unity, which requires their supports to overlap throughout the surface.
Adaptive density control maintains this condition as the geometry evolves, in the spirit of the densification of 3DGS~\cite{kerbl2023gaussian}.
It has two parts: a criterion that identifies the surfels whose current size and position fail to represent the surface detail, and a refinement rule that decides what to do with them.

\paragraph{Densification criterion.}
\label{sec:grad_accum}
In 3DGS~\cite{kerbl2023gaussian}, the criterion is the magnitude of the gradient of the loss with respect to the projected 2D center $\boldsymbol{\mu}_i$ of each primitive, averaged over the training views,
\begin{equation}
  \bar{g}_i = \frac{1}{N_i}\sum_{j=1}^{N_i}
  \left\|\frac{\partial \mathcal{L}_j}{\partial \boldsymbol{\mu}_i}\right\|_2\,,
  \label{eq:3dgs_grad}
\end{equation}
where $N_i$ counts the views in which primitive $i$ contributes to the image.
In 2DGS~\cite{huang20242d}, whose primitives are not projected through an affine approximation, this gradient is approximated by projecting the gradient with respect to the 3D center onto the image plane, which mixes depth and lateral motion.
With the T-matrix, the 2D center is available in closed form: it is the projection of the center column of $\mathbf{T}$, $(p_x, p_y) = (T_{1,z}/T_{4,z},\; T_{2,z}/T_{4,z})$, where $T_{i,z}$ denotes the entry of row $i$ in the column of the center.
Holding $T_{4,z}$ fixed, a screen-space translation of the surfel changes only $T_{1,z}$ and $T_{2,z}$, so that the per-pixel contributions to the screen-space positional gradient are
\begin{equation}
  g_x = \frac{\partial\mathcal{L}}{\partial T_{1,z}} \cdot T_{4,z}\,,\qquad
  g_y = \frac{\partial\mathcal{L}}{\partial T_{2,z}} \cdot T_{4,z}\,.
  \label{eq:ss_grad}
\end{equation}
Following AbsGS~\cite{ye2024absgs} and Yu et al.~\cite{yu2024gaussian}, we accumulate the norms of these per-pixel contributions rather than the norm of their sum, which avoids the cancellation of gradients of opposite signs from different pixels of the same surfel; the score of a surfel is the mean of these accumulated norms over the views in which it contributes, as in \cref{eq:3dgs_grad}.

\paragraph{Density-aware splitting.}
\label{sec:splitting}
A score above a threshold, $\bar{g} \ge \tau_g$, signals that the local reconstruction is inadequate, but does not determine the remedy.
3DGS clones a primitive with a high score if it is small and splits it if it is large, with a size threshold defined as a fixed fraction of the scene extent.
Such a global threshold does not adapt to the local sampling density, and cannot enforce the spatially varying overlap condition of the coverage model.
We instead compare the scales of each surfel with the local sample spacing.
For each surfel, a spatial index over the surfel centers gives the mean distance $\bar{d}$ to its orientation-aware nearest neighbors, \ie, the neighbors whose normals are compatible with its own, so that the estimate reflects the local surface rather than unrelated geometry at a different depth.
In the scattered-data approximation framework~\cite{wendland2004scattered}, the support of a reconstruction kernel should scale with this spacing.
A surfel is therefore split when $\bar{g} \ge \tau_g$ and its larger scale exceeds the local spacing, $\max(s_1, s_2) > \bar{d}$, which targets over-reconstructed surfels whose support is too wide for the local point distribution.

A split surfel is replaced by two children at $\mathbf{c} \pm \tfrac12 s_{\max}\,\hat{\mathbf{t}}_{\max}$, along its normalized major axis $\hat{\mathbf{t}}_{\max}$.
The scale along the split axis is multiplied by $0.7$, which leaves enough overlap between the siblings to keep the reconstructed signal continuous, while the other scale is preserved; all other attributes are inherited from the parent.
This deterministic two-way split keeps both children coplanar with the parent, preserving the local surface, and increases the number of surfels gradually.

\paragraph{Pruning.}
\label{sec:pruning}
Densification alone is not sufficient: the coverage model is equally sensitive to surfels that should not exist, since any surfel that accumulates weight at a pixel contributes to the coverage and the color of that pixel.
We therefore remove surfels with four criteria, each targeting a failure mode that the photometric loss alone does not penalize.
\begin{itemize}
  \item \emph{Visibility.} At the end of each pass over the training views, surfels whose accumulated screen-space gradient (\cref{eq:ss_grad}) is zero are removed. Such a surfel was never reached by the backward pass, because it was persistently occluded, outside all camera frustums, or rejected by the cutoff test at every pixel, and therefore carries no information about the scene.
  \item \emph{Isolation.} A surfel with no neighbor within its kernel support has no overlap with the rest of the point cloud, and cannot take part in the associative overlap of kernels that defines a surface layer (\cref{sec:depth_grouping}).
  \item \emph{Silhouette.} When masks are available, surfels that project outside the foreground in a sufficient fraction of the training views in which they are visible are removed as floaters.
  \item \emph{Back-facing occluders.} A surfel whose normal points away from the camera, yet which appears in the front surface layer, is geometrically inconsistent, since it would be invisible on a closed surface. Such surfels arise from the orientation ambiguity of the tangent frame.
\end{itemize}
Each criterion removes at most a fixed fraction of the surfels per invocation, so that a misfiring criterion degrades the reconstruction gracefully rather than collapsing it, and all criteria are disabled once densification ends, so that the final iterations optimize a fixed set of surfels.

\subsection{Losses}
\label{sec:supp_losses}

The photometric losses $\mathcal{L}^{\text{sh}}$ and $\mathcal{L}^{\text{pbr}}$ of the SH and physically-based images, the silhouette loss $\mathcal{L}_{\text{sil}}$, and the depth consolidation loss $\mathcal{L}_{\text{cons}}$ (\cref{eq:cons_loss}) are defined in \cref{sec:loss_function}.
We detail here the remaining regularizers.

\paragraph{Normal consistency.}
The normal consistency loss penalizes the angular deviation between the normal of a surfel and those of its orientation-aware nearest neighbors,
\begin{equation}
  \mathcal{L}_{\text{nc}} = \frac{\sum_{i}\sum_{j \in \mathcal{N}_i} w_{ij}\,(1 - \mathbf{n}_i \cdot \mathbf{n}_j)}{\sum_{i}\sum_{j \in \mathcal{N}_i} w_{ij}}\,,
  \label{eq:nc_loss}
\end{equation}
where $\mathcal{N}_i$ is the set of nearest neighbors of surfel~$i$ with a compatible normal orientation, and $w_{ij} = \exp(-d_{ij}^2 / 2\sigma_i^2)$ decays with the distance $d_{ij}$ between the centers of surfels $i$ and $j$, with $\sigma_i$ derived from the local sample spacing.
The neighbor index is rebuilt periodically and cached between updates to amortize its cost.

\paragraph{Depth-normal consistency.}
Being local, $\mathcal{L}_{\text{nc}}$ cannot detect a patch of surfels that is smooth but globally misoriented.
The depth-normal consistency loss catches such patches by comparing, in screen space, the rendered normal map $\hat{\mathbf{n}}$ with the normals $\mathbf{n}^{\text{depth}}$ of the rendered depth,
\begin{equation}
  \mathcal{L}_{\text{nd}} = \frac{1}{|\mathcal{V}|}\sum_{p \in \mathcal{V}}
    \bigl(1 - \hat{\mathbf{n}}_p \cdot \mathbf{n}^{\text{depth}}_p\bigr)\,,
  \label{eq:nd_loss}
\end{equation}
evaluated on a validity set $\mathcal{V}$ that excludes weakly covered pixels, pixels adjacent to depth discontinuities, and, when masks are available, background pixels.

\paragraph{Illumination and material priors.}
Three further priors, in the spirit of those of R3DG~\cite{gao2023relightable} and IRGS~\cite{gu2024IRGS}, regularize the under-constrained separation of material and illumination.
A white-light prior $\mathcal{L}_{\text{light}}$ penalizes the per-channel deviation of the mean incident direct radiance from its achromatic mean, expressing that a colored observation is more likely to come from a colored material than from a colored illuminant.
A total-variation term $\mathcal{L}_{\text{tv}}$ on the environment map suppresses the high-frequency texture that a free per-texel light would otherwise use to fit view-dependent residuals.
Edge-aware smoothness terms $\mathcal{L}_{\rho}$ and $\mathcal{L}_{r}$ on the rendered albedo and roughness maps, weighted by the gradients of the input images, encode that material changes are sparse and tend to coincide with image edges.
These two terms are applied with small weights: each can push the material in only one direction, whereas the sign of the error it would correct varies across scenes, so that their effect is scene-dependent.

\paragraph{Total losses.}
With scalar weights $\lambda$, the geometry stage and the inverse-rendering stage minimize
\begin{align}
  \mathcal{L}^{(1)} &= \mathcal{L}^{\text{sh}}
    + \lambda_{\text{pbr}}\,\mathcal{L}^{\text{pbr}}
    + \lambda_{\text{sil}}\,\mathcal{L}_{\text{sil}}
    + \lambda_{\text{cons}}\,\mathcal{L}_{\text{cons}} \nonumber\\
  &\quad + \lambda_{\text{nc}}\,\mathcal{L}_{\text{nc}}
    + \lambda_{\text{nd}}\,\mathcal{L}_{\text{nd}}
    + \lambda_{\ell}\,\mathcal{L}_{\text{light}}\,,
  \label{eq:total_loss}\\
  \mathcal{L}^{(2)} &= \mathcal{L}^{\text{pbr}}
    + \lambda_{\text{sh}}\,\mathcal{L}^{\text{sh}}
    + \lambda_{\ell}\,\mathcal{L}_{\text{light}}
    + \lambda_{\text{tv}}\,\mathcal{L}_{\text{tv}} \nonumber\\
  &\quad + \lambda_{\rho}\,\mathcal{L}_{\rho}
    + \lambda_{r}\,\mathcal{L}_{r}\,.
  \label{eq:total_loss2}
\end{align}
The SH loss remains in $\mathcal{L}^{(2)}$ because the SH radiance supplies the indirect illumination of the secondary rays (\cref{sec:supp_exclusion}).

\subsection{Training Schedule}
\label{sec:training_schedule}

Training proceeds in the two stages of \cref{sec:optimization}, optimized with Adam~\cite{kingma2014adam}.

\paragraph{Stage 1: geometry.}
The geometry stage runs for 30k iterations and proceeds through two phases.
In the \emph{warm-up} phase, the photometric losses of both images and the silhouette loss are the only active terms, which lets the surfels deform toward the observed images under the sparse initial coverage.
In the \emph{densification} phase, adaptive density control (\cref{sec:density_control}) is enabled, splitting and pruning the surfels at regular intervals, and the geometric regularizers $\mathcal{L}_{\text{cons}}$, $\mathcal{L}_{\text{nc}}$, and $\mathcal{L}_{\text{nd}}$ are activated.
Throughout the stage, the learning rate of the surfel positions decays exponentially and is scaled by the spatial extent of the scene, and the SH degree is annealed upward.

\paragraph{Stage 2: materials and illumination.}
At the stage boundary, the geometry is frozen, the albedo, roughness, metallic, and environment map are re-initialized, the BVH is built, and the secondary rays are traced once (\cref{sec:supp_caching}).
The stage then runs for 20k iterations, minimizing $\mathcal{L}^{(2)}$ (\cref{eq:total_loss2}) with the Monte Carlo shading of \cref{sec:supp_mc}.

\section{Experimental Setup}
\label{sec:supp_setup}
\label{sec:experimental_setup}

\paragraph{Training configuration.}
On both datasets, models are trained on a single NVIDIA RTX A6000 GPU, with 30k iterations in the geometry stage, 20k iterations in the inverse-rendering stage, and $S = 64$ Monte Carlo samples per surfel.
Stanford-ORB~\cite{kuang2023stanfordorb} models are trained at $512{\times}512$ resolution from the HDR input images.
The learning rates follow the conventions of 3DGS-based methods.

\paragraph{Surfel budgets.}
Densification and pruning bring the Stanford-ORB models from about 100K initial surfels to about 200K surfels on average.
On TensoIR Synthetic~\cite{jin2023tensoir}, the optimized models contain about 300K surfels on average, but Ficus and Lego require about 500K surfels each.
Both objects contain thin structures that are observed from both sides, such as the leaves of the Ficus and the thin parts of the Lego.
Our renderer does not support double-sided shading: a surfel is back-face culled, and therefore seen and shaded only from the side its normal points to.
A thin sheet must therefore be represented by two layers of surfels with opposite orientations, one for each side, which increases the number of surfels it requires.
The spherical initialization of the ablation study (\cref{tab:tensoir_ablation}) starts from 200K surfels, which densification and pruning bring to a comparable count.

\paragraph{Baselines.}
On Stanford-ORB (\cref{tab:benchmark}), we use the results provided by the benchmark~\cite{kuang2023stanfordorb} or by the respective authors~\cite{zhou2025rtr}, except for IRGS~\cite{gu2024IRGS} and NVDiffRecMC~\cite{hasselgren2022shape}, which we train ourselves with their official implementations.
Some methods do not report some metrics because they are not applicable or not available.
On TensoIR (\cref{tab:tensoir}), we likewise train IRGS and NVDiffRecMC ourselves; all other results, including those of NVDiffRec and the runtimes, are taken from GS-IR~\cite{liang2024gs}.
For the qualitative comparisons, the images of IRGS and NVDiffRecMC are rendered from the models we trained.

\paragraph{Geometry evaluation.}
The Chamfer distances of \cref{tab:benchmark} and the meshes of \cref{fig:geometry} are obtained by reconstructing a mesh from the optimized surfels with NKSR~\cite{huang2023nksr}, with the settings given in \cref{sec:surface_extraction}.

\paragraph{Fixed-geometry configuration.}
For the configuration 3DSS\textsuperscript{\textdagger} of \cref{tab:benchmark}, we sample surfels uniformly on the ground-truth laser scans with the anisotropic procedure of \cref{sec:mesh_sampling}.
The scans contain about 200K faces, and we sample on average 5 surfels per face, which yields about 1M surfels per object.
Only the shading and the illumination are optimized.

\section{Performance Analysis}
\label{sec:performance}

We analyze the computational cost of 3DSS along two axes: the end-to-end training time, and the throughput of the rasterizer in isolation.

\paragraph{Training time.}
The full two-stage optimization of 3DSS (30k + 20k iterations) completes in under one hour per scene on average on a single NVIDIA RTX A6000 GPU.
This is on par with GS-IR and NVDiffRec, and faster than NVDiffRecMC~\cite{hasselgren2022shape}, which needs up to two hours because its ray-traced shading adds to the cost of every iteration (\cref{tab:tensoir}).
3DSS is slightly slower than IRGS~\cite{gu2024IRGS}, which we attribute to our use of forward rather than deferred shading.
IRGS shades the rasterized attribute buffers once per pixel, whereas we shade every surfel before rasterization (\cref{sec:forward_shading}), so that the shading cost grows with the number of surfels rather than with the number of pixels, and is also spent on surfels that are occluded or whose radiance is averaged with that of their neighbors within a layer.
The secondary-ray trace of the second stage is not a significant part of this cost, since it runs only once (\cref{sec:supp_caching}).
In the geometry stage, 3DSS shares the per-iteration cost structure of 3DGS-based approaches: projection, tile-based binning, sorted rasterization, and back-propagation through the surfel primitives, and we therefore expect training times comparable to those of other Gaussian and surfel methods at similar primitive counts and resolutions.

\begin{figure}[t]
  \centering
   \includegraphics[width=\columnwidth]{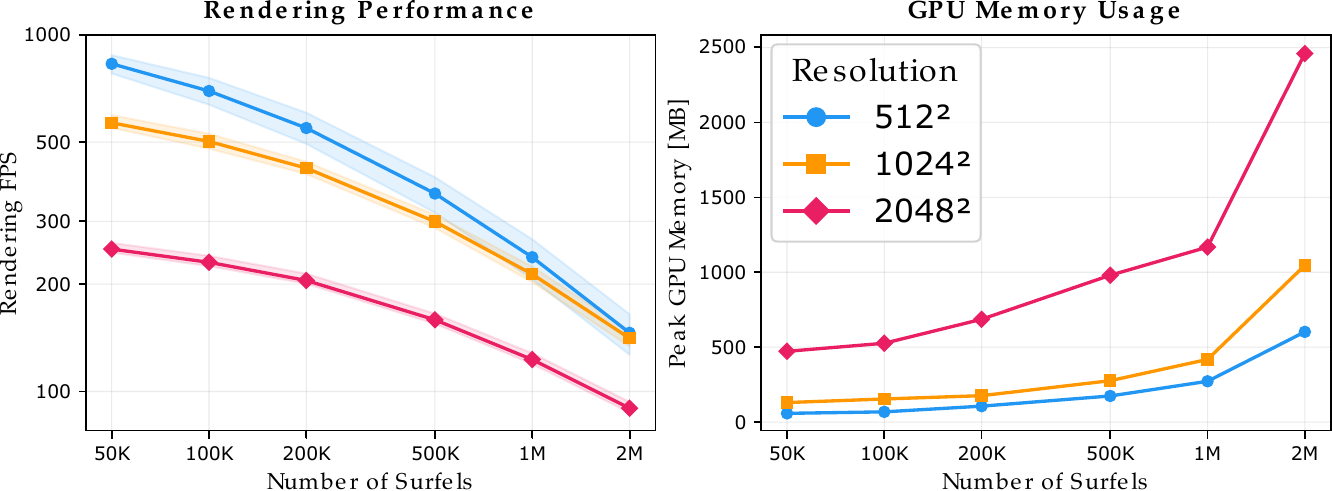}
  \caption{\textbf{Rendering performance of the 3DSS rasterizer.}
  Rendering speed (left, log scale) and peak GPU memory (right) as a function of the surfel count, for three output resolutions, measured on a point-sampled Stanford Bunny that nearly fills the viewport.
  Timings isolate the rasterizer by rendering the albedo only.
  Each data point is the mean over a 600-frame orbital trajectory; shaded bands indicate $\pm$ one standard deviation.}
  \label{fig:performance}
\end{figure}

\paragraph{Rendering throughput.}
To characterize the rasterizer independently of the shading model, we benchmark it on a point-sampled Stanford Bunny.
The bunny mesh is converted to surfels with the procedure of \cref{sec:mesh_sampling}, and we vary both the surfel count ($50$K, $100$K, $200$K, $500$K, $1$M, and $2$M) and the output resolution ($512{\times}512$, $1024{\times}1024$, and $2048{\times}2048$).
The camera is placed so that the object nearly fills the viewport, which minimizes the fraction of empty pixels and ensures that the measurements reflect sustained rasterization load rather than early termination on background tiles.
Each configuration is rendered along a 600-frame orbital trajectory, preceded by a 200-frame warm-up that excludes the transient costs of kernel compilation and memory allocation.
The configuration is \emph{albedo-only}: the per-surfel albedo is rendered without evaluating any shading function.

\Cref{fig:performance} reports the results.
Across all tested configurations, the rasterizer runs above the $60$\,fps interactive threshold: about $830$\,fps at $512{\times}512$ with $50$K surfels, and still about $90$\,fps at $2048{\times}2048$ with $2$M surfels.
The frame rate decreases sublinearly with both the pixel count and the surfel count: a 16-fold increase in pixel count, from $512{\times}512$ to $2048{\times}2048$, lowers it by a factor of about three up to $200$K surfels, and a 40-fold increase in surfel count, from $50$K to $2$M, by a factor of three to six depending on the resolution.
Memory consumption remains moderate: at the operating points of our experiments ($200$K to $300$K surfels, $512{\times}512$ to $1024{\times}1024$), the peak GPU memory stays below 300\,MB.
It exceeds 1\,GB only with $1$M surfels or more at $2048{\times}2048$, where it reaches about 2.5\,GB with $2$M surfels, and marginally with $2$M surfels at $1024{\times}1024$.

\section{Interoperability with Meshes}
\label{sec:supp_meshes}

\Cref{fig:teaser} illustrates that 3DSS converts in both directions between surfels and triangle meshes.
From surfels to meshes, the optimized surfels form an oriented point cloud from which a mesh is reconstructed (\cref{sec:surface_extraction}), and their materials are then baked into the mesh as vertex attributes and texture maps (\cref{sec:baking}).
From meshes to surfels, 3DSS renders oriented point clouds sampled from triangle meshes (\cref{sec:mesh_sampling}).

\subsection{Surface Extraction}
\label{sec:surface_extraction}

Each surfel carries a position $\mathbf{c}$ and a unit normal $\mathbf{n}$, so that the optimized surfels form an \emph{oriented point cloud} in the classical sense of the surface reconstruction literature~\cite{hoppe1992surface}, without any auxiliary extraction procedure.
Algorithms that reconstruct surfaces from oriented points, such as Neural Kernel Surface Reconstruction (NKSR)~\cite{huang2023nksr}, which we use in our experiments, or Screened Poisson Surface Reconstruction (SPSR)~\cite{kazhdan2013screened}, apply directly to the output of the optimization and yield a triangle mesh.
This contrasts with volumetric representations such as 3DGS~\cite{kerbl2023gaussian}, where recovering an explicit surface requires indirect procedures, typically the TSDF fusion of rendered depth maps~\cite{huang20242d} or the coupling to an auxiliary mesh~\cite{guedon2024sugar}, which add pipeline stages and scene-dependent parameters.
In our formulation, the point cloud \emph{is} the surface representation.

\paragraph{Reconstruction settings.}
The surfel centers alone sample the surface only at the density of the surfels, and convey neither their extent nor their anisotropy.
We therefore augment each surfel with 16 additional oriented points, sampled on its support disk within twice its scales and carrying its normal, so that the reconstruction sees the surface that the surfels actually cover.
We reconstruct with the pretrained \texttt{snet} model of NKSR at a voxel size of $0.004$ in scene units, keep the largest connected component, and close the remaining holes.
These settings are used for all the meshes and Chamfer distances that we report.

\paragraph{Interpolating and approximating regimes.}
The choice of reconstruction algorithm interacts with the Shepard normalization of the rasterizer.
Where the surfel kernels overlap enough to approximate a partition of unity, the normalized accumulation acts as a \emph{kernel interpolant}: the reconstructed signal passes through or near the sample values, and the optimized surfel centers settle close to the target surface.
Where the partition-of-unity condition is not met, for instance at surface boundaries or in regions of rapid geometric variation, the normalization behaves closer to a \emph{kernel regression}, or Nadaraya--Watson estimator~\cite{nadaraya1964estimating,watson1964smooth}, which produces a smoothed approximation of the underlying function rather than an interpolant.
In this regime, the surfel centers need not lie exactly on the true surface, but at positions that minimize the photometric objective under the smoothing bias of the estimator.
Methods that assume that the input points interpolate the surface, such as SPSR with a high interpolation weight, may then produce artifacts in under-sampled regions, whereas approximating methods with a smoothness prior, such as NKSR, absorb this residual positional bias.
SPSR remains effective with a reduced interpolation weight (its \texttt{pointWeight} parameter), which relaxes the interpolation constraint and lets the Poisson solve regularize the surface.

\subsection{Material Baking}
\label{sec:baking}

The reconstructed mesh carries the geometry of the surfels but none of their materials.
We transfer the materials in two steps (\cref{fig:teaser}, right): first to the vertices of the mesh, which yields per-vertex attributes, and then to texture maps over a UV atlas, which yields an asset that standard rendering engines and glTF pipelines can load.
Neither step refits anything to the images: every baked value is an interpolation of the optimized surfel attributes.

\paragraph{Transfer kernel.}
The material field of 3DSS is defined not by the surfel centers alone but by their reconstruction kernels, so a vertex should read it through the same kernel that the rasterizer integrates.
For a vertex $\mathbf{x}$ and a surfel with center $\mathbf{c}$, normalized tangent axes $\hat{\mathbf{t}}_u, \hat{\mathbf{t}}_v$, normal $\mathbf{n}$, and scales $(s_1, s_2)$, we express the offset $\mathbf{x} - \mathbf{c}$ in the frame of the surfel,
\begin{equation}
  u = \frac{(\mathbf{x} - \mathbf{c})\cdot\hat{\mathbf{t}}_u}{s_1}\,,\quad
  v = \frac{(\mathbf{x} - \mathbf{c})\cdot\hat{\mathbf{t}}_v}{s_2}\,,\quad
  h = (\mathbf{x} - \mathbf{c})\cdot\mathbf{n}\,,
  \label{eq:bake_coords}
\end{equation}
and weight the surfel by
\begin{equation}
  w = \exp\!\Bigl(-\frac{u^2 + v^2}{2\,\sigma_t^2}\Bigr)\,
      \exp\!\Bigl(-\frac{h^2}{2\,(\sigma_n \bar{s})^2}\Bigr)\,,
  \label{eq:bake_kernel}
\end{equation}
where $\bar{s}$ is the mean scale of the surfel.
With $\sigma_t = 1$, the tangential factor is exactly the object-space kernel of \cref{eq:kernel}, so that an anisotropic surfel reaches the mesh along its major axis and not across it.
The normal factor accounts for the offset between the reconstructed surface and the surfel planes, which can reach a voxel of the reconstruction; we use $\sigma_n = 2$.
It also resolves thin structures geometrically, without a back-face test: a surfel on the opposite side of a thin feature lies far from the vertex along its own normal, and receives a negligible weight.

Each vertex gathers its $k = 96$ nearest surfel centers, and its albedo, roughness, and metallic are the Shepard-normalized averages (\cref{eq:shepard}) of their attributes under the weights of \cref{eq:bake_kernel}.
The neighbor count $k$ only needs to be large enough that the kernel, and not the size of the neighborhood, sets the extent of the average; we verify that the weight of the farthest neighbor stays below 1\% of the largest weight at the median vertex.
Beyond this point, a larger $k$ only smooths the baked material.
The surfel normals are averaged in the same way after being flipped into the hemisphere of the vertex normal, and renormalized.
The sign alignment matters, because misoriented surfel normals tend to occur in coherent patches rather than as isolated outliers, and two neighbors on opposite sides of a thin sheet would otherwise cancel into a near-zero vector whose normalization is meaningless.

One might instead interpolate from the augmented oriented points to which the mesh is fitted (\cref{sec:surface_extraction}), with an isotropic kernel in Euclidean distance.
These points are strongly correlated, since each surfel contributes 17 of them: the nearest points of a vertex come from only a few surfels, and the bake then averages far fewer independent materials than the rasterizer, which integrates every surfel whose support covers a pixel.
In our experiments, this alternative produced a visibly noisier albedo than the splatted material render, and a lower agreement with it than the kernel of \cref{eq:bake_kernel}.

\paragraph{Two normal fields.}
The baked mesh carries two normal fields, which serve different purposes.
The \emph{geometric normals}, area-weighted averages of the face normals at each vertex, describe the reconstructed surface and are used for shading.
The \emph{baked normals} are the interpolated surfel normals, \ie, the orientation field recovered by the optimization, and are the ones to compare with the normal maps rendered by 3DSS.
The two fields differ slightly, by about $5^\circ$ at the median vertex, and the tangent-space normal map of the textured asset encodes exactly this difference.

\paragraph{Textured asset.}
To produce a UV-mapped asset, we first decimate the mesh to 200K faces by quadric edge collapse~\cite{garland1997surface}, which bounds the cost of the UV parameterization; the per-vertex bake above always runs on the full-resolution mesh.
The materials are transferred to the decimated vertices directly from the surfels with \cref{eq:bake_kernel}, rather than from the full-resolution bake, so that decimation never compounds two interpolations.
A UV atlas is computed and packed with xatlas\footnote{\url{https://github.com/jpcy/xatlas}} at a target resolution of $2048{\times}2048$ texels; since the packer enlarges the atlas when the charts do not fit, the textures follow the packed size, which preserves the padding between charts.
The per-vertex attributes are then rasterized into the atlas and stored as 8-bit albedo, roughness, metallic, and normal maps.
The normal map is expressed in tangent space: per-face tangents are computed from the UV derivatives, orthogonalized against the geometric vertex normal by Gram--Schmidt, and given a bitangent sign from the UV winding of each face, so that mirrored charts do not flip the map.
Since this frame is built from the geometric normals, the map stores only the deviation of the surfel orientation field from the mesh surface, and remains valid under any rigid transformation of the asset.
Finally, every texel outside the charts receives the value of its nearest texel inside a chart, computed with a Euclidean distance transform.
This dilation is not cosmetic: bilinear texture fetches near a chart boundary read texels just outside the chart, which would otherwise appear as dark seams in the albedo and as glossy seams in the roughness.

\begin{table}[t]
\centering
\footnotesize
\setlength{\tabcolsep}{3.5pt}
\caption{\textbf{Fidelity of material baking on Stanford-ORB.} PSNR (dB) within the shared silhouette, and silhouette IoU. Top: per-vertex bake of the full-resolution mesh against the 3DSS surfel model, both shaded with direct lighting only. Bottom: exported textured asset, rendered from disk, against the per-vertex bake.}
\label{tab:baking}
\begin{tabular}{@{}ll ccccc@{}}
\toprule
Object & View & IoU & Albedo & Rough. & Normal & Shaded \\
\midrule
\multicolumn{7}{@{}l}{\emph{Per-vertex bake vs.\ 3DSS}} \\
Gnome (sc.~3) & train & 0.9830 & 33.67 & 31.34 & 39.38 & 35.99 \\
Gnome (sc.~3) & test  & 0.9844 & 29.25 & 28.39 & 34.72 & 34.45 \\
Gnome (sc.~7) & train & 0.9814 & 27.87 & 27.73 & 33.58 & 28.41 \\
Cactus (sc.~5) & train & 0.9759 & 30.54 & 27.93 & 35.28 & 34.23 \\
\midrule
\multicolumn{7}{@{}l}{\emph{Textured asset vs.\ per-vertex bake}} \\
Gnome (sc.~3) & train & 0.9995 & 43.16 & 37.85 & 49.07 & 36.61 \\
Gnome (sc.~3) & test  & 0.9997 & 40.08 & 36.43 & 39.10 & 37.82 \\
Gnome (sc.~7) & train & 0.9995 & 38.04 & 35.74 & 43.48 & 31.79 \\
Cactus (sc.~5) & train & 0.9996 & 44.41 & 42.06 & 46.57 & 41.02 \\
\bottomrule
\end{tabular}
\end{table}

\paragraph{Validation.}
We verify the bake by rendering the mesh with the shading model of 3DSS.
The mesh is rasterized with \textsc{nvdiffrast}~\cite{laine2020modular} into per-pixel positions, normals, albedo, and roughness, which are shaded with the same Monte Carlo estimator (\cref{eq:mc_shading}), environment map, and sample directions as the surfels.
Since the mesh has no secondary-ray tracer, we compare it with 3DSS rendered with direct lighting only, \ie, with $V = 1$ and $L_{\text{ind}} = 0$ in \cref{eq:incident}; the two renders then differ only in the primitive that carries the surface, and every difference between them is attributable to the reconstruction and the bake.
\Cref{tab:baking} (top) reports their agreement within the shared silhouette, on three captures chosen to span smooth and thin geometry.
Rendering the mesh twice with independent sample sets gives a self-agreement of 41.1\,dB, above all values of the top block, which therefore reflect differences in geometry and material rather than Monte Carlo noise.
The remaining gap is largest where the surface reconstruction merges thin features that the surfels resolve separately, rather than in the transfer itself.
\Cref{tab:baking} (bottom) isolates what the export adds to the per-vertex bake, \ie, decimation, UV parameterization, and 8-bit quantization: the material maps agree at 36 to 49\,dB PSNR on all views, the test view included.

\subsection{Surfels from Meshes}
\label{sec:mesh_sampling}

Conversely, 3DSS renders surfels sampled from a triangle mesh, which is how the fixed-geometry configuration 3DSS\textsuperscript{\textdagger} is built (\cref{sec:supp_setup}) and how the benchmark of \cref{sec:performance} is set up.
Point positions and normals are given directly by the sampling of the faces.
The remaining parameters, namely the tangent vectors and the scales, are estimated from the local neighborhood of each sample, with one of two strategies.
\begin{enumerate}[label=(\roman*)]
  \item \emph{Isotropic scaling.}
  The tangent vectors are an arbitrary orthonormal frame of the tangent plane, and the scale is set isotropically to the mean distance to the orientation-aware nearest neighbors, as for the initialization (\cref{sec:supp_init}) and the splitting criterion (\cref{sec:density_control}), with the same spatial index.
  This is computationally efficient, but an isotropic kernel conforms poorly to sharp geometric edges, where it produces minor visual artifacts.
  \item \emph{Anisotropic scaling.}
  The neighbors that share a compatible normal orientation are projected onto the tangent plane of the sample, and a 2D principal component analysis is performed on the projected coordinates.
  The eigenvectors of the resulting $2\!\times\!2$ covariance matrix define the tangent directions, and its eigenvalues determine anisotropic scales that adapt the kernel to the local point distribution.
  This removes the edge artifacts of isotropic kernels, at the cost of additional computation and memory for the eigendecomposition.
\end{enumerate}
The resulting scale pairs are converted to the bounded parameterization of \cref{sec:parameterization}, which clamps any anisotropy beyond $\kappa$.
We use the anisotropic variant for 3DSS\textsuperscript{\textdagger} (\cref{tab:benchmark}) for its quality; the isotropic variant is a drop-in replacement when speed matters.

\section{Ablation Study: Extended Discussion}
\label{sec:supp_ablation}

This appendix extends our discussion of the optimization ablations (\cref{sec:ablation_components} and \cref{tab:tensoir_ablation}).
So that the comparisons reflect the quality of the surfel distribution rather than differences in model capacity, the densification and pruning parameters of every configuration are calibrated so that the converged models reach a per-scene surfel budget comparable to that of the full method.

\subsection{Adaptive Density Control}
\label{sec:ablation_densification}

We disable splitting (\cref{sec:density_control}) while keeping all other components, including pruning, unchanged.
Without the ability to refine the surfel distribution, under-sampled regions retain the coarse coverage of the initial point cloud, which limits both the geometric detail and the material resolution.
The degradation is largest in novel view synthesis, which indicates that what is lost is mostly high-frequency appearance and thin structures that the initial surfels cannot represent.
Adaptive refinement is therefore what allows the optimization to exceed the resolution of its initialization.

\subsection{Regularization Losses}
\label{sec:ablation_losses}

We disable in turn each of three regularizers of the geometry stage: $\mathcal{L}_{\mathrm{cons}}$, $\mathcal{L}_{\mathrm{nc}}$, and $\mathcal{L}_{\mathrm{sil}}$.
Their effect on TensoIR is smaller than that of adaptive density control, as expected for clean synthetic captures with accurate masks and camera poses, but each of them improves every metric of the study.

\paragraph{Depth consolidation ($\mathcal{L}_{\mathrm{cons}}$).}
Removing $\mathcal{L}_{\mathrm{cons}}$ leaves relighting essentially unchanged and slightly degrades novel view synthesis, albedo, and normals.
Its main effect is on the geometry of the surfel cloud, which the image metrics capture poorly (\cref{fig:ddist_ablation}): without it, multi-layer compositing allows the optimizer to distribute surfels across several depth layers, and the point cloud becomes thicker and more dispersed along the viewing direction.
By penalizing both the spread of the compositing weight across layers and the depth variance within each layer, $\mathcal{L}_{\mathrm{cons}}$ enforces the thinness that a surface representation should have, and on which the oriented-point reconstruction of \cref{sec:surface_extraction} relies.

\paragraph{Normal consistency ($\mathcal{L}_{\mathrm{nc}}$).}
Disabling $\mathcal{L}_{\mathrm{nc}}$ barely affects novel view synthesis, but it is the most damaging of the three loss ablations for albedo and relighting.
This is what one should expect when the normals are left unconstrained: each surfel orients itself to minimize the photometric residual of the pixels it covers, and the training views can still be reproduced, because a misoriented surfel can compensate with a wrong albedo or roughness.
The compensation is invisible under the training illumination and fails under any other.
The normal consistency loss thus removes a degree of freedom that the photometric loss cannot constrain, in exchange for normals that generalize to unseen illumination.

\paragraph{Silhouette ($\mathcal{L}_{\mathrm{sil}}$).}
Removing the silhouette loss mainly degrades novel view synthesis, while albedo, relighting, and normals change little.
Without explicit silhouette supervision, the surfels receive only a weak signal to cover the object boundary, and stray surfels appear early in training that the subsequent densification phase must correct.
The silhouette loss thus improves the surfel distribution along object boundaries, but its effect on the material and geometry estimates is small once the geometry has converged.

\subsection{Sensitivity to Initialization}
\label{sec:initialization_exp}

The default initialization unprojects the depth maps predicted by Depth Anything~3~\cite{lin2026depth} (\cref{sec:supp_init}).
It introduces a dependency on an external model that may be unavailable or unreliable, for instance in domains where pretrained depth estimators perform poorly.
We therefore evaluate an initialization from 200K surfels sampled uniformly on a sphere enclosing the object, from which the geometry must be recovered by the photometric and silhouette losses alone.
This configuration falls behind the full model on all metrics, most clearly in novel view synthesis and normal accuracy.

The monocular depth prior places the initial surfels near the object surface from the first iteration.
This helps in regions where the shading alone cannot disambiguate the geometry, such as concavities visible from few viewpoints or surfaces whose appearance is dominated by interreflections.
The predicted depth maps contain noise and bias, but on TensoIR these errors are corrected during optimization and do not outweigh the benefit of the prior.

The spherical initialization inherits no such errors, but all surfels must migrate to the surface by gradient descent, and in geometrically ambiguous regions, the shading model and the regularizers are the only source of signal.
In practice, the optimization then falls into local minima with worse geometry, especially on thin structures such as the branches and leaves of the Ficus: surfels initialized far from them rarely receive a consistent gradient pulling them onto the surface, and the final model settles on an incomplete or oversmoothed configuration.
Part of this sensitivity stems from the fact that the gradient does not account for changes in the assignment of surfels to surface layers (\cref{sec:backward}).

\section{Additional Qualitative Results}
\label{sec:supp_qualitative}

\subsection{Relighting on TensoIR Synthetic}
\label{sec:supp_tensoir_relight}
Complementing the relighting comparison of \cref{fig:tensoir_relight}, \cref{fig:supp_tensoir_relight} shows additional relit test views of the four TensoIR Synthetic scenes (Armadillo, Ficus, Hotdog, and Lego) under the held-out \emph{night}, \emph{bridge}, \emph{forest}, and \emph{city} environment maps, compared with NVDiffRecMC~\cite{hasselgren2022shape}, IRGS~\cite{gu2024IRGS}, and the ground truth.

\begin{figure*}[t]
  \centering
  \setlength{\tabcolsep}{0.5pt}
  \renewcommand{\arraystretch}{0.6}
  \newcommand{\RLw}{0.115\textwidth}
  \newcommand{\RLdir}{Figures/Comparisons/TensoIR/relight_supp}
  \newcommand{\RLpair}[3]{%
    \figimg[width=\RLw]{\RLdir/#2/#1.jpg} &
    \figimg[width=\RLw]{\RLdir/#3/#1.jpg}}
  \newcommand{\RLrow}[2]{%
    \RLpair{nvdiffrecmc}{#1}{#2} & \RLpair{irgs}{#1}{#2} &
    \RLpair{ours}{#1}{#2} & \RLpair{gt}{#1}{#2}}
  \newcommand{\RLlabel}[1]{\smash{\rotatebox{90}{\makebox[\dimexpr\RLw*2+1pt\relax][c]{\small #1}}}}
  \begin{tabular}{@{}c cc@{\hskip 4pt}cc@{\hskip 4pt}cc@{\hskip 4pt}cc@{}}
  & \multicolumn{2}{c}{\small NVDiffRecMC} & \multicolumn{2}{c}{\small IRGS} &
    \multicolumn{2}{c}{\small \textbf{3DSS (Ours)}} & \multicolumn{2}{c}{\small GT} \\[2pt]
  & \RLrow{armadillo_test_062_night}{armadillo_test_079_bridge} \\[1pt]
  \RLlabel{Armadillo} &
    \RLrow{armadillo_test_058_forest}{armadillo_test_083_city} \\[3pt]
  & \RLrow{ficus_test_163_night}{ficus_test_159_bridge} \\[1pt]
  \RLlabel{Ficus} &
    \RLrow{ficus_test_150_forest}{ficus_test_174_city} \\[3pt]
  & \RLrow{hotdog_test_098_night}{hotdog_test_174_bridge} \\[1pt]
  \RLlabel{Hotdog} &
    \RLrow{hotdog_test_002_forest}{hotdog_test_065_city} \\[3pt]
  & \RLrow{lego_test_195_night}{lego_test_183_bridge} \\[1pt]
  \RLlabel{Lego} &
    \RLrow{lego_test_190_forest}{lego_test_167_city} \\
  \end{tabular}
  \caption{\textbf{Additional relighting comparison on TensoIR Synthetic.}
  Relighting results of NVDiffRecMC~\cite{hasselgren2022shape}, IRGS~\cite{gu2024IRGS}, and 3DSS against the ground truth for Armadillo, Ficus, Hotdog, and Lego.
  Within each $2{\times}2$ block, the four tiles show different test views relit under the held-out \emph{night} (top left), \emph{bridge} (top right), \emph{forest} (bottom left), and \emph{city} (bottom right) environment maps.
  Relit objects are composited onto the corresponding environment map.}
  \label{fig:supp_tensoir_relight}
\end{figure*}

\subsection{Stanford-ORB}
\label{sec:supp_orb}
Complementing \cref{fig:main_comparison}, \cref{fig:supp_orb_1,fig:supp_orb_2,fig:supp_orb_3} show six additional Stanford-ORB objects (Ball, Car, Chips, Gnome, Salt, and Teapot), with novel view synthesis, albedo, normals, and relighting for NVDiffRecMC~\cite{hasselgren2022shape}, IRGS~\cite{gu2024IRGS}, 3DSS, and the ground truth.
Each object is shown from a held-out view of its training scene; the relighting tile shows another view of the object under the illumination of a different capture.
Albedo and relighting are scale-aligned, normals are shown in camera space, and tiles are cropped to the object.

\newlength{\orbsh}
\newcommand{\ORBSimg}[3]{\figimg[height=\orbsh]{Figures/Comparisons/ORB_supp/#1/#2_#3.png}}
\newcommand{\ORBSrow}[4]{%
  \rotatebox{90}{\makebox[\orbsh][c]{\scriptsize #4}} &
  \ORBSimg{#1}{nvs}{#3} & \ORBSimg{#1}{albedo}{#3} & \ORBSimg{#1}{normal}{#3} & \ORBSimg{#1}{relight}{#3} &
  \ORBSimg{#2}{nvs}{#3} & \ORBSimg{#2}{albedo}{#3} & \ORBSimg{#2}{normal}{#3} & \ORBSimg{#2}{relight}{#3}}
\newcommand{\ORBStable}[5]{%
  \setlength{\orbsh}{\dimexpr(\textwidth-24pt)*1000/#5\relax}%
  \centering
  \setlength{\tabcolsep}{0.5pt}\renewcommand{\arraystretch}{0.6}%
  \begin{tabular}{@{}c cccc@{\hskip 6pt}cccc@{}}
  & \multicolumn{4}{c}{\small #2} & \multicolumn{4}{c}{\small #4} \\
  \cmidrule(lr){2-5}\cmidrule(lr){6-9}
  & {\scriptsize NVS} & {\scriptsize Albedo} & {\scriptsize Normal} & {\scriptsize Relight}
  & {\scriptsize NVS} & {\scriptsize Albedo} & {\scriptsize Normal} & {\scriptsize Relight} \\[1pt]
  \ORBSrow{#1}{#3}{nvdiffrecmc}{NVDiffRecMC} \\
  \ORBSrow{#1}{#3}{irgs}{IRGS} \\
  \ORBSrow{#1}{#3}{3dss}{\textbf{3DSS (Ours)}} \\
  \ORBSrow{#1}{#3}{gt}{GT} \\
  \end{tabular}}

\begin{figure*}[t]
  \ORBStable{ball}{Ball}{car}{Car}{8006}
  \caption{\textbf{Additional Stanford-ORB comparison (1/3).} Ball (\emph{scene002}, relit under \emph{scene003}) and Car (\emph{scene002}, relit under \emph{scene006}). Albedo and relighting are scale-aligned.}
  \label{fig:supp_orb_1}
\end{figure*}

\begin{figure*}[t]
  \ORBStable{chips}{Chips}{gnome}{Gnome}{7604}
  \caption{\textbf{Additional Stanford-ORB comparison (2/3).} Chips (\emph{scene002}, relit under \emph{scene004}) and Gnome (\emph{scene003}, relit under \emph{scene005}). Albedo and relighting are scale-aligned.}
  \label{fig:supp_orb_2}
\end{figure*}

\begin{figure*}[t]
  \ORBStable{salt}{Salt}{teapot}{Teapot}{7433}
  \caption{\textbf{Additional Stanford-ORB comparison (3/3).} Salt (\emph{scene004}, relit under \emph{scene005}) and Teapot (\emph{scene006}, relit under \emph{scene001}). Albedo and relighting are scale-aligned.}
  \label{fig:supp_orb_3}
\end{figure*}

\end{document}